\documentclass{aa}
\usepackage[varg]{txfonts}
\pdfoutput=1
\usepackage{graphicx}
\usepackage{newtxtext,newtxmath}
\usepackage[colorlinks=true, linkcolor=blue, citecolor=blue, filecolor=blue, urlcolor=blue]{hyperref}

\newcommand{\MSUN}{\rm {M}_{\odot}}

\newcommand{\MTWOC}{{M}_{\rm 200c, host}}

\usepackage{color}
\usepackage[normalem]{ulem}

\begin{document}

\title{X-ray-inferred kinematics of the core intracluster medium in Perseus-like clusters:\\
Insights from the TNG-Cluster simulation}
\titlerunning{X-ray kinematics of Perseus-like clusters in TNG-Cluster}

\author{Nhut Truong\inst{1,2}\thanks{E-mail: ntruong@umbc.edu}
\and Annalisa Pillepich\inst{3}
\and Dylan Nelson\inst{4}
\and Irina Zhuravleva\inst{5}
\and Wonki Lee\inst{6,7}
\and \\Mohammadreza Ayromlou\inst{4}
\and Katrin Lehle\inst{4}
}

\institute{NASA Goddard Space Flight Center, Greenbelt, MD 20771, USA \label{1}
\and Center for Space Sciences and Technology, University of Maryland, 1000 Hilltop Circle, Baltimore, MD 21250, USA \label{2}
\and Max-Planck-Institut f{\"u}r Astronomie, K{\"o}nigstuhl 17, 69117 Heidelberg, Germany \label{3}
\and Universit\"{a}t Heidelberg, Zentrum f\"{u}r Astronomie, ITA, Albert-Ueberle-Str. 2, 69120 Heidelberg, Germany \label{4}
\and Department of Astronomy and Astrophysics, The University of Chicago, Chicago, IL 60637, USA \label{5}
\and Yonsei University, Department of Astronomy, Seoul, Republic of Korea \label{6}
\and Harvard-Smithsonian Center for Astrophysics, 60 Garden St., Cambridge, MA 02138, USA \label{7}
}

\date{}

\abstract{
The intracluster medium (ICM) of galaxy clusters encodes the impact of the physical processes that shape these massive halos, including feedback from central supermassive black holes (SMBHs). In this study, we examine the gas thermodynamics, kinematics, and the effects of SMBH feedback on the core of Perseus-like galaxy clusters with a new simulation suite: TNG-Cluster. We first make a selection of simulated clusters similar to Perseus based on the total mass and inner ICM properties, such as their cool-core nature. We identify 30 Perseus-like systems among the 352 TNG-Cluster halos at $z=0$. Many exhibit thermodynamical profiles and X-ray morphologies with disturbed features such as ripples, bubbles, and shock fronts that are qualitatively similar to X-ray observations of Perseus. To study observable gas motions, we generate XRISM mock X-ray observations and conduct a spectral analysis of the synthetic data. In agreement with existing Hitomi measurements, TNG-Cluster predicts subsonic gas turbulence in the central regions of Perseus-like clusters, with a typical line-of-sight velocity dispersion of 200 km/s. This implies that turbulent pressure contributes $< 10\%$ to the dominant thermal pressure. In TNG-Cluster, such low (inferred) values of ICM velocity dispersion coexist with high-velocity outflows and bulk motions of relatively small amounts of super-virial hot gas, moving up to thousands of km/s. However, detecting these outflows in observations may prove challenging due to their anisotropic nature and projection effects. Driven by SMBH feedback, such outflows are responsible for many morphological disturbances in the X-ray maps of cluster cores. They also increase both the inferred and intrinsic ICM velocity dispersion. This effect is somewhat stronger when velocity dispersion is measured from higher-energy lines. 
}
\maketitle


\section{Introduction}
\label{sec:intro}

In the standard $\Lambda$CDM model, large-scale structure assembles hierarchically, with progressively larger and more massive objects forming from the smooth accretion of matter and mergers with smaller halos. In this scenario, galaxy clusters form last, at the latest epochs in cosmic history, and are the largest virialized systems in the Universe (see \citealt{kravtsov.borgani.2012}). 

Gravity
is the dominant driver in such a high-mass regime (virial mass ${\rm\sim 10^{15}M_\odot}$), and dictates the  physical state and properties
of the clusters (\citealt{kaiser.1986}). However, it is expected that non-gravitational physical processes such as active galactic nuclei (AGN) feedback, which results from gas accretion onto the central supermassive black hole (SMBH), may also play an important role in shaping and modulating the physical state of the matter components in clusters, as suggested by numerous numerical works (e.g. \citealt{puchwein.etal.2008,short.etal.2010,lebrun.etal.2014, planelles.etal.2014,rasia.etal.2015, li.etal.2015,truong.etal.2018,pop.etal.2022}). Studying the impact of SMBH feedback is not only intriguing from a physical standpoint but is also essential for using clusters as cosmological probes (\citealt{voit.2005, vikhlinin.etal.2009,allen.evrard.mantz.2011}). 

One possible avenue to probe the effects of SMBH feedback is via the X-ray analysis of the intracluster medium (ICM), which is a diffuse, hot, and fully ionised plasma (with a temperature of about $10^{7-8}$ K) that fills the cluster halos. ICM surrounds the cluster member galaxies and the central brightest galaxy (\citealt{bohringer.werner.2011}). The central SMBH deposits its energy into the ICM. Therefore, the ICM thermodynamics and kinematics are affected by and encode information about this feedback (see \citealt{hlavacek-Larrondo.etal.2022, donahue.voit.2022} for a review). The average global properties of the ICM on large spatial scales (e.g. within the virial radius of 1-2 Mpc) are chiefly dictated by the cluster mass via scaling relations, as predicted by the self-similar model (\citealt{kaiser.1986,giodini.etal.2013}). On the other hand, at smaller scales (e.g. within the cluster core and closer to the central AGN engine at $\lesssim$ 100 kpc from its centre) the ICM is likely more susceptible to the feedback. Indeed AGN feedback is widely considered as a plausible heating source that prevents cooling flows in cluster cores \citep{churazov.etal.2000, McNamara.etal.2000,fabian.2002}.

The Perseus cluster is one of the most studied clusters across the energy and wavelength spectrum. It is a system of ${\rm M_{200c} \sim 10^{14.7-15}\rm{M}_\odot}$ in total mass\footnote{${\rm M_{200c}}$ is defined as the total mass contained within a radius of ${\rm R_{200c}}$, where the average density is 200 times the critical density of the Universe.} located at a distance of $\sim 73$ Mpc. Over the past decade or so, X-ray observations with Chandra and other X-ray telescopes have uncovered an X-ray surface brightness map that is rich in structures and spatial fluctuations
 \citep{fabian.etal.2011}, and this map shows that the core of Perseus is characterised by a disturbed morphology compared to non-cool-core clusters. Furthermore, these observations have revealed an abundance of phenomena in the ICM of the Perseus core including cavities or bubbles, shock fronts, cold fronts, X-ray filaments, and many other features that are yet to be identified. The feature-rich X-ray morphology of Perseus is thought to originate from a diverse array of processes, including the feedback activities driven by the central AGN in NGC1275 \citep{fabian.etal.2011} and the gas sloshing presumably caused by a minor merger with a subcluster, among others (e.g. \citealt{zuhone.etal.2013}). AGN feedback likely dominates the disturbances within the innermost $\sim60$ kpc region (e.g. \citealt{heinrich.etal.2021}), while outside the 60 kpc core the ICM perturbations may be increasingly driven by gas sloshing (e.g. \citealt{walker.etal.2018}). 

Regardless of their physical origin, the perturbations in the X-ray maps suggest the existence of turbulent motions in the central regions of Perseus. Turbulent motions are believed to be an important heating channel that helps maintain cool-core clusters such as Perseus. Based on the analysis of X-ray surface brightness fluctuations, turbulent dissipation is sufficient to offset the cooling rate in the core of Perseus  (\citealt{zhuravleva.etal.2014}). 

The first direct measurement of gas motions in a galaxy cluster indeed targeted Perseus and was conducted by Hitomi \citep{hitomi.collaboration.2016}. Using its advanced, microcalorimeter-based spectroscopic capabilities, Hitomi was able to measure gas velocities via the shifting and broadening of emission lines due to the Doppler effect as well as via the resonant scattering effect \citep{hitomi.collaboration.2018}. These brief observations of Perseus offer a glimpse into the kinematic structures of the gas within the central $\sim100$ kpc region of the cluster. The observation reveals almost uniform motion of the hot ICM in the Perseus core, with a line-of-sight velocity dispersion of about 160 km/s, implying that the gas-motion-induced pressure is only $4\%$ of the thermal pressure. Within the uncertainties, the measured velocities are consistent with the analysis of the surface brightness fluctuations (\citealt{zhuravleva.etal.2018}).

The JAXA-led X-ray mission XRISM (X-ray Imaging and Spectroscopy Mission)\footnote{\url{https://xrism.isas.jaxa.jp/en/}} was recently launched. The mission will provide Hitomi-like spectroscopic data not only for Perseus but also for other massive clusters in the nearby Universe.  

Theoretically, it has been challenging to unravel the feature-rich X-ray morphology of Perseus as well as its kinematic structures, given that those features may emerge from a complex interplay among various multi-scale and coupled processes. We therefore turn to numerical simulations to understand the X-ray observations of Perseus. Several simulations have focused on the Perseus cluster \citep[e.g.][]{zuhone.etal.2018,bourne.sijacki.2017,hillel.soker.2017,lau.etal.2017, heinrich.etal.2021}. However, it is worth noting that most of these studies rely on idealised simulations of a single halo. Typically, these simulations are tailored to reproduce only a limited number of observable ICM properties, and of the Perseus cluster alone. As a consequence, the findings from these studies may not be applicable to the broader cosmological context of cluster formation. On the other hand, a small number of investigations employed cosmological simulations \citep[e.g.][]{lau.etal.2017}, although these simulations generally have lower numerical resolution and simplified AGN feedback models, limiting their ability to reproduce the same level of complexity observed in the Perseus X-ray observations.

In the present paper, we complement previous theoretical and numerical studies and explore, quantify, and provide interpretation of current and future kinematical measurements of the core of Perseus as well as other clusters. Our work is based on a new suite of cosmological magnetohydrodynamical simulations named TNG-Cluster (\citealt{nelson.etal.2023}, \textcolor{blue}{Pillepich
et al. in prep}). Leveraging this novel simulation suite, we produced and analysed X-ray synthetic data derived from XRISM mock observations of a sample of simulated cool-core clusters in the Perseus mass range. The main goals of this study are fourfold: (i) to demonstrate that the outcome of TNG-Cluster is sufficiently realistic and diverse to return systems whose physical properties are similar to those of the real Perseus cluster; (ii) to quantify the level of turbulent or chaotic motions in the ICM predicted by state-of-the-art full cosmological models of clusters by conducting an end-to-end analysis of the data as if observed with XRISM; (iii) to quantify the impact of SMBH feedback on driving gas motions in the central regions of clusters  across multiple possible realisations of cool-core galaxy clusters; and (iv) to provide guidance for the interpretation of current and future measurements of the velocity structure of the ICM via X-ray observations.

This is one of a series of papers where we showcase TNG-Cluster, and present its first scientific results: the overall simulation suite, its technical details and the basic properties of the clusters are presented by \cite{nelson.etal.2023}. In a companion paper, \cite{ayromlou.etal.2023} quantifies ICM gas kinematics, from cluster cores to their outskirts. \cite{lehle.etal.2023} study cool-core versus non-cool-core clusters and their buildup over time; \cite{lee.etal.2023} identify merging systems and the diversity, morphology, and physical properties of radio relics; and \cite{rohr.etal.2023} discuss how cluster satellites can retain their circumgalactic media, and also provide information on its observability and implications.
 
This paper is arranged as follows. In Section~\ref{sec:methods} we introduce the TNG-Cluster simulation and the definition and computation of simulated quantities, and we outline the detailed pipeline of mock X-ray analysis. The main results from TNG-Cluster are given in Section \ref{sec:results}, where we present the selection and thermodynamical properties of a subsample of cool-core clusters in the Perseus mass range (Section~\ref{sec:selection}); we also quantify the mock X-ray measurements of the gas motions in the cores of the simulated Perseus-like clusters (Section~\ref{sec:Xray_measurements}). In Section~\ref{sec:intepretation}, we first quantify possible systematic uncertainties of both X-ray-based and intrinsic measurements of the gas velocity dispersions (Section~\ref{sec:systematics}), before characterising the dependence of the measured velocity dispersion on cluster properties and SMBH activity (Section~{\ref{sec:m200_mbh_dependence}). In Sections \ref{sec:outflows} and \ref{sec:sigma_ratio}, we explore the relationship between velocity dispersion and high-velocity outflows and their impact on the thermodynamics of the core ICM. Finally, we summarise our results and outline our conclusions in Section~\ref{sec:summary}.  


\section{Methods: Simulated clusters and X-ray mock observations}
\label{sec:methods}

\subsection{The TNG-Cluster simulation suite}

TNG-Cluster\footnote{\url{www.tng-project.org/cluster}} is a cosmological magneto-hydrodynamical simulation suite of massive galaxy clusters (\citealt{nelson.etal.2023} and Pillepich et al. in prep}). It is a spin-off project of IllustrisTNG \citep[TNG hereafter;][]{springel.etal.2018,pillepich.etal.2018b,nelson.etal.2018,naiman.etal.2018,marinacci.etal.2018}, which is a series of cosmological simulations of galaxy formation and evolution with various simulated volumes and resolutions: TNG50, TNG100, and TNG300. 

TNG-Cluster was developed to improve the sampling and statistics of the highest halo-mass end: halo masses of  ${\rm M_{200c}\gtrsim10^{14.5}\,\MSUN}$. It comprises $352$ zoom-in simulations that are drawn from a 1Gpc parent box. The simulations are performed with the \textsc{AREPO} code \citep{springel.2010} and has mass and spatial resolutions equivalent to the highest resolution realisation of TNG300, namely $1.1\times10^{7}\rm{M}_\odot$ in baryon mass, ${\rm 6.1\times10^{7}M_\odot}$ in DM mass, and a Plummer equivalent gravitational softening of $\sim1.5$ kpc at $z=0$ for the collisionless components. TNG-Cluster is based on the same ${\rm \Lambda}$CDM cosmological model as TNG, with cosmological parameters consistent with Planck analysis of the primordial microwave background (\citealt{planck.collaboration.2016}): ${\rm \Omega_m=0.3089}$, ${\rm\ \Omega_\Lambda=0.6911}$, ${\rm \Omega_b=0.0486}$, ${\rm \sigma_8=0.8159}$, ${\rm n_s=0.9667}$, and ${\rm H_0=67.74\ km/s/Mpc}$.

For galaxy astrophysics, TNG-Cluster is fully based on the IllustrisTNG model \citep{pillepich.etal.2018b}, which has been well tested against numerous observations of galaxies. This model implements various processes that are relevant to forming and evolving galaxies, including: (i) radiative cooling from primordial/metal-line and heating from a UV/X-ray background radiation field, (ii) star formation, (iii) evolution of stellar populations and metal enrichment produced by type  Ia and II supernovae and AGB stars, accounting for the production of nine tracked elements: H, He, C, N, O, Ne, Mg, Si, and Fe, in addition to Europium; (iv) supernovae-driven stellar feedback; and (v) the seeding, growth, and dual-mode (thermal and kinetic) feedback of SMBHs. 

Of relevance for this paper, in the TNG model and so also in TNG-Cluster, feedback energy from SMBHs is injected into the surrounding environment in the form of either thermal energy or kinetic energy \citep{weinberger.etal.2017, pillepich.etal.2018}, in addition to a radiative-like mode that modulates the cooling of the halo gas \citep{vogelsberger.etal.2013}. SMBH feedback is crucial in order to reproduce realistic properties of massive galaxies such as the galaxy luminosity function at low redshift \citep{pillepich.etal.2018b}, the galaxy colour bimodality \citep{nelson.etal.2018}, and the quenched fractions \citep{donnari.etal.2021b}. Furthermore, in the TNG model, SMBH feedback has a profound impact on the gaseous halos around galaxies and groups and clusters of galaxies, where its effects manifest as various phenomena also at the Milky-Way and Andromeda galaxy-mass scale: an X-ray luminosity dichotomy between star-forming and quiescent galaxies \citep{truong.etal.2020}, X-ray eROSITA-like bubbles in Milky-Way like discy galaxies \citep{pillepich.etal.2021}, and overall an anisotropic distribution of the thermodynamics, chemistry, and magnetic fields  of the circumgalactic medium (CGM) *\citep{truong.etal.2021b, peroux.etal.2020, ramesh.etal.2023d}. Furthermore, the TNG model also reproduces X-ray and Sunyaev-Zeldovich (SZ) scaling relations in galaxies, groups, and clusters, showing broad agreement with observations (\citealt{pop.etal.2022}).

In summary, the TNG-Cluster simulations are built upon a well-tested galaxy formation model, in which the SMBH feedback component is not only crucial for reproducing realistic samples of galaxies but also has the ability to realise a wide range of observationally supported phenomena imprinted on the hot gaseous atmospheres around them.

\subsection{Definition and computation of simulated quantities}
\label{sec:definition}

In this work, we exclusively focus on the 352 zoom-in galaxy clusters at redshift $z=0$ targeted and simulated within TNG-Cluster. These span the mass range log$_{\rm 10}(\MTWOC / \MSUN) = 14.3 - 15.4$ at $z=0$ and include all matter components, from cold dark matter to stars. In the following, we focus on the SMBHs and the gas properties of these simulated clusters by considering all gas cells belonging to each cluster according to the friends-of-friends halo-finding algorithm (\citealt{davis.etal.1985}). The cluster properties we use throughout are:

\begin{itemize}
    \item {\it Core status.} Following the companion paper by \cite{lehle.etal.2023}, we categorise TNG-Cluster systems, that is, we classify their core status, based on the central cooling time of the gas, which is defined as 
    \begin{equation}
    t_{\rm cool}=\frac{3}{2}\frac{(n_e+n_i)k_B T}{n_e n_i \Lambda}, \label{eq:tcool}
    \end{equation}
    where $n_e$ ($n_i$) is the electron (ion) number density of a gas cell, $T$ is its temperature, $\Lambda$ is the cooling function, and $k_B$ is the Boltzmann constant. The cooling time of a cluster is computed as a mass-weighted average of gas cells within the central region of radius $0.012R_{500c}$. Generally, the core status of a cluster is determined based on its central cooling time in comparison with the age of the Universe at the formation epoch. In practice, assuming the average formation redshift of $z=0$ clusters is $z=1$, a cluster is defined as a non cool-core if its cooling time exceeds 7.7 Gyr (e.g. \citealt{ODeal.etal.2008,McDonald.etal.2013}). Specifically, we classify the core status as follows:
    \begin{itemize}
        \item {\bf Strong cool-core (SCC)}: ${\rm t_{cool}<1\ Gyr}$.
        \item {\bf Weak cool-core (WCC)}: ${\rm 1\ Gyr\leq t_{cool}<7.7\ Gyr}$.
        \item {\bf Non cool-core (NCC)}: ${\rm t_{cool}\geq 7.7\ Gyr}$.
    \end{itemize}
    \item {\it Gas kinematics.} The mean velocity ($v_z$) and velocity dispersion ($\sigma_z$) of the gas along an arbitrary direction ---for example along the z-axis of the simulation box--- are defined as:
    \begin{equation}
        v_{z, w}=\frac{\sum_{i}w_i v_{z,i}}{\sum_i w_i}, \label{eq:mean_vel}
    \end{equation}
    \begin{equation}
        \sigma_{z,w}^2 = \frac{\sum_i w_i v_{z,i}^2}{\sum_i w_i}-v_{z,w}^2, \label{eq:mean_veldis}
    \end{equation}
    where $w$ is either gas mass (mass-weighted, mw) or X-ray emission (emission-weighted, ew), and the $i$-index runs over the gas cells belonging to the selected region. Specifically, the X-ray emission of each gas cell is computed in the [6.0-8.0] keV energy range based on the gas density, temperature, and metallicity and by assuming an APEC emitting model using the XSPEC package \citep{smith.etal.2001}. We use these estimates of the intrinsic velocity dispersion of the gas as a benchmark for that obtained by X-ray spectral fitting, as described below.
\end{itemize}

\subsection{Mock X-ray analysis}
\label{sec:mock_Xray}

In this paper, we aim to create synthetic XRISM Resolve X-ray data from the TNG-Cluster simulation output and from those we aim to extract thermodynamical and kinematic measurements (i.e. velocity dispersion) of the ICM as carried out in observational studies. For this purpose, we constructed an end-to-end pipeline for the analysis of synthetic X-ray spectra, which consists of two main steps.

In the first step, X-ray emission from the diffuse gas is derived based on the intrinsic gas thermodynamics, including density, temperature, and elemental abundances, as returned by the simulation. We use the PyXSIM package \citep{zuhone.etal.2014}, which was developed from the PHOX algorithm (\citealt{biffi.etal.2012, biffi.etal.2013}), to create samples of mock X-ray photons. To model the X-ray emission of the ICM, we use the APEC model with atomic data taken from AtomDB \citep[version 3.0.8;][]{smith.etal.2001}, which assumes that the modelled plasma is optically thin and in collisional ionisation equilibrium (CIE). Mock X-ray photons are subjected to the Doppler effect due to both thermal and gas motions. Furthermore, the mock photons are subjected to galactic absorption, which is described by the `wabs' model. Finally, we simulate the interaction between the incoming photons and the XRISM Resolve detector with the SOXS package \citep{zuhone.etal.2023}, employing the XRISM Resolve instrumental responses (e.g. ARF and RMF files\footnote{\url{https://heasarc.gsfc.nasa.gov/docs/xrism/proposals/}}), assuming a spectral resolution of 5 eV. This first step produces an event file that registers mock observed photons from simulated clusters, and resembles those produced for real X-ray observations. We note that in our X-ray mock observations, we do not include instrumental and cosmic X-ray backgrounds, as these backgrounds are expected to be significantly smaller than the signals coming from the cluster cores in the [6.4-8.0] keV band where the spectral data are used for fitting. This energy band is selected following the approach used in the Hitomi work on Perseus (\citealt{hitomi.collaboration.2016}), where this range encompasses the most prominent lines of the spectrum, including the Fe XXV He-${\alpha}$ complex, Fe XXVI Ly$\alpha$, and Fe XXV He$\beta$. These lines are used for the measurements of gas motions. In addition, the use of these high-energy lines is advantageous because of their high resolving power (${\rm E/dE\sim1300-1600}$), and enhances the precision of the velocity dispersion measurements.

Once the event files are created, we extract spectra using dmextract from the CIAO package. The spectra are then fitted to a single component `bapec' model using the XSPEC package \citep{smith.etal.2001}, which accounts for both thermally and velocity-based line broadening. As a result, the fit returns a number of best-fitting quantities, such as redshift (i.e. line shift), normalisation (gas density), temperature, metallicity, and Gaussian sigma for the velocity broadening. 

\begin{figure*}
    \centering
 \includegraphics[width=0.69\textwidth]{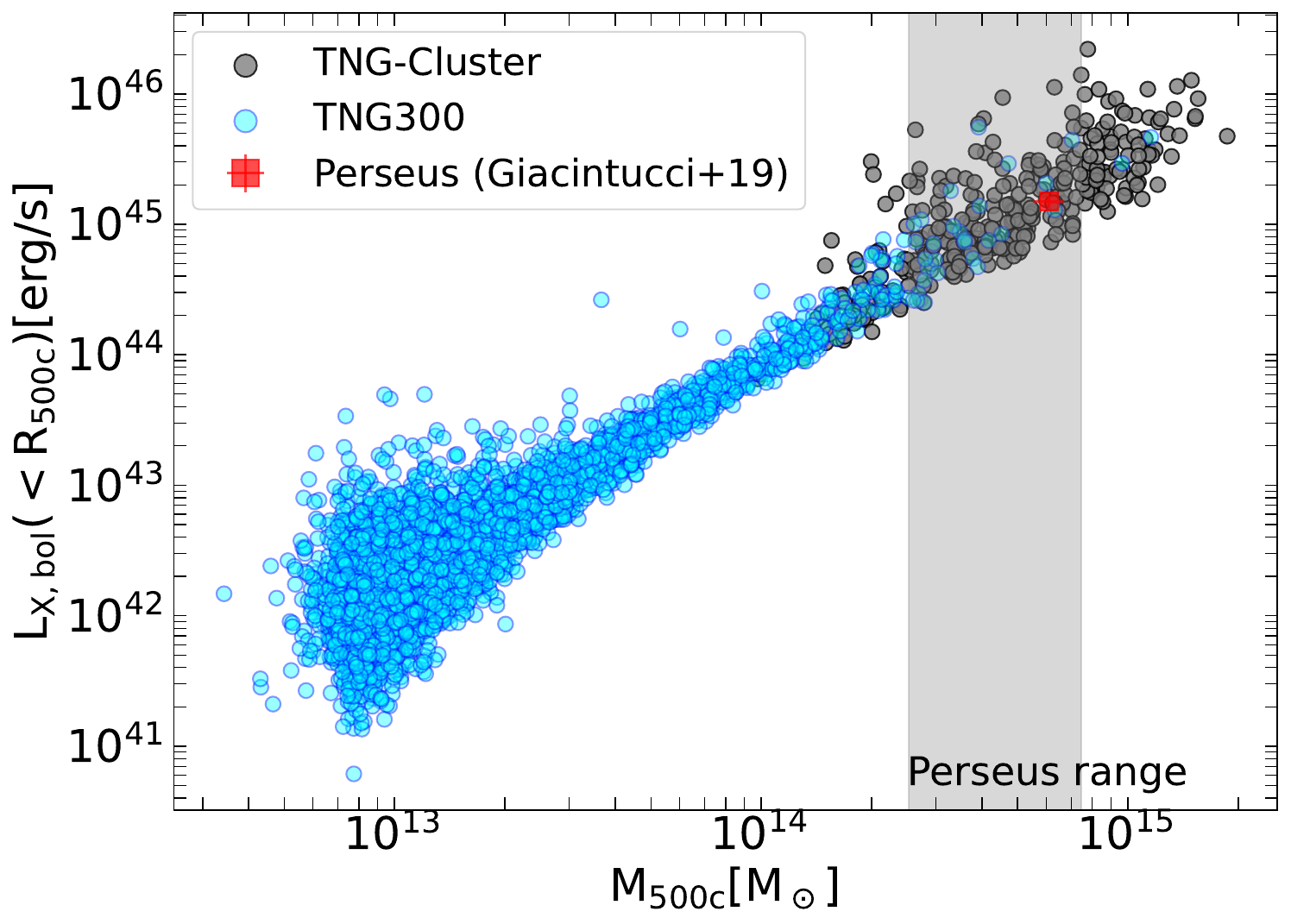}
 \includegraphics[width=0.99\textwidth]{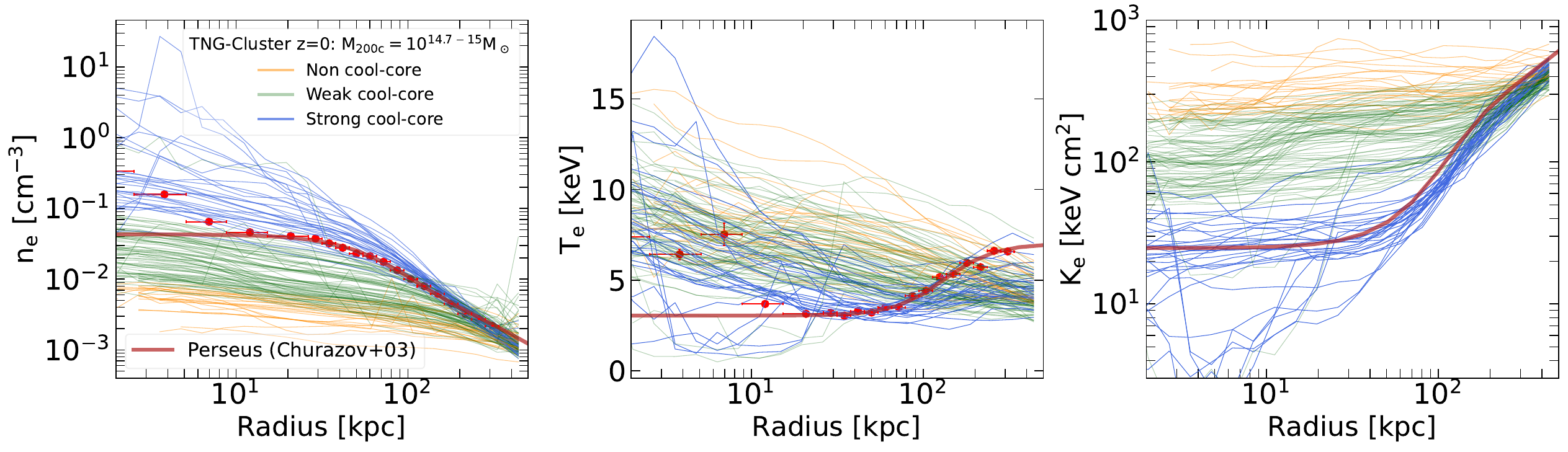}
    \caption{{\bf Selection and properties of Perseus-like clusters from the TNG-Cluster simulation suite.} {\it Top:} Distribution of clusters in the ${\rm L_{X}-M_{500c}}$ plane at $z=0$, from TNG-Cluster (grey) and TNG300 (blue), in comparison with the Chandra observed data of Perseus from \cite{giacintucci.etal.2019}. In the Perseus mass range, ${\rm M_{200c}=10^{14.7-15.0}\rm{M}_\odot}$ (or ${\rm M_{500c}=10^{14.4-14.87}M_\odot}$), there are 135 TNG-Cluster halos. {\it Bottom: }Intrinsic thermodynamical profiles of the selected clusters within the Perseus mass range in comparison ---i.e. at face value, without replicating the observational measurements--- with the XMM-Newton observed data and analytical approximations from \citealt{churazov.etal.2003} (red). From left to right, we show the 3D radial profiles of electron number density, temperature, and entropy ($K_{\rm e}\equiv k_{\rm B} T_{\rm e}/n_{\rm e}^{2/3}$). The selected clusters are further classified according their core status: non cool-core (orange), weak cool-core (green), and strong cool-core (blue). All profiles are computed as averages weighted by bolometric X-ray emission. In this paper, we focus on Perseus-like clusters, namely those that are strong cool cores (blue profiles, based on central cooling time) and in the Perseus mass range.}
    \label{fig:selection}
\end{figure*}

Mock X-ray observations are created for each simulated cluster viewed in a random orientation (the z-axis of the TNG-Cluster box), which is referred to as the line of sight (LoS). In principle, we could greatly expand the sample by considering each simulated system from multiple projections and at slightly different points in time: this is not needed to support the current scientific results, but could be done in future studies.


\section{Core kinematics results  with TNG-Cluster}
\label{sec:results}

\subsection{Perseus-like systems from the TNG-Cluster suite}
\label{sec:selection}

In this paper, we focus on Perseus-like clusters in order to compare the outcome of TNG-Cluster to Hitomi measurements and to make predictions for upcoming results with XRISM.
Among the 352 clusters at $z=0$, we identify a subsample of simulated clusters with properties that closely resemble those observed in the Perseus cluster. In particular, the selection is based on two criteria, which respectively characterise the clusters from a global and an inner-region perspective: 
\begin{enumerate}
    \item Halo mass: We consider systems with ${\rm M_{200c}=10^{14.7-15}\,\rm{M}_\odot}$ based on optical and X-ray inferences of Perseus' total mass (e.g. \citealt{simionescu.etal.2011, Aguerri.etal.2020}). This halo mass range corresponds to ${\rm M_{500c}=10^{14.40-14.87}M_\odot}$.
    \item Core properties: We select only strong cool-core systems, which have a central cooling time of below 1 Gyr (see Section~\ref{sec:definition}). This selection is in line with X-ray observations of Perseus, which indicate that it is a cool core characterised by a notably short cooling time at its centre ($\sim0.3$ Gyr, \citealt{sanders.etal.2004, sanders.etal.2007}).
\end{enumerate}

TNG-Cluster produces 30 strong cool-core systems within the selected mass range at the $z=0$ snapshot. We focus on these for the rest of the paper and refer to them as Perseus-like clusters. Prior to exploring the kinematics of these selected clusters, we will first examine their thermodynamics in relation to existing observations of the Perseus cluster.

\begin{figure*}
    \centering
        \includegraphics[width=0.99\textwidth]{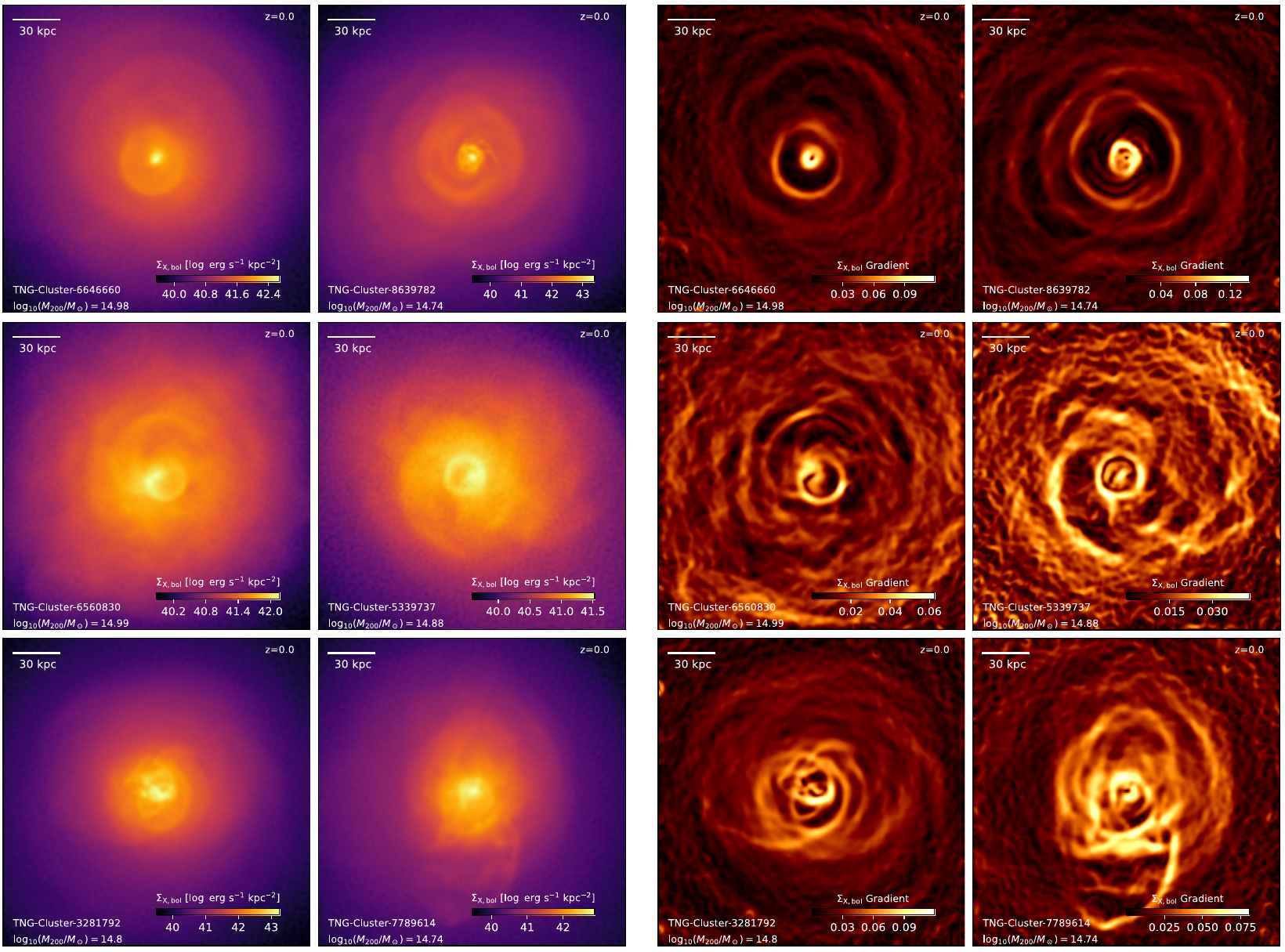}
    \caption{{\bf X-ray surface brightness ({\it left column}) and spatial gradient ({\it right column}) maps of a subset of Perseus-like systems from the TNG-Cluster simulation.} The gradient maps are made by applying a Gaussian gradient magnitude filter with a smoothing length ${\rm \sigma=2}$ pixels ($2$ kpc) to the surface brightness maps on the left column. The maps encompass a region of 200 kpc per side with a projection depth of $2R_{\rm 200c}$. It is noted that the color scale varies arcoss the different clusters. Halos that exhibit closely analogous X-ray morphologies are grouped within the same row. As observed in the real Perseus, ripples, cavities or bubbles, and shock fronts are clearly visible in our simulated Perseus-like clusters at $z=0$.} 
    \label{fig:xray_maps}
\end{figure*}

In Fig.~\ref{fig:selection} we quantify the thermodynamical properties of the Perseus-like systems from TNG-Cluster, and compare them with pre-existing X-ray observations of Perseus. The top panel displays the luminosity--mass scaling relation for all TNG-Cluster halos, as well as for TNG300, with the observed data of Perseus provided by Chandra (\citealt{giacintucci.etal.2019}). TNG-Cluster extends the ${\rm L_X-M_{500c}}$ relation ---where ${\rm L_X}$ is measured within ${\rm R_{500c}}$--- into the high-mass regime (${\rm M_{500c}\geq 5\times10^{14}M_\odot}$), where TNG300, because of its smaller simulated volume, does not capture the full scatter of this relation. The observed luminosity of Perseus falls well within the selected mass range of the Perseus-like subsample from our simulation. 

In the bottom subpanels of Fig.~\ref{fig:selection}, we show the spherically symmetric radial profiles of the mass-selected clusters for electron density, electron temperature, and entropy. All are computed as 3D emission-weighted profiles, taking emission as X-ray bolometric luminosity, and therefore, they are intrinsic quantities measured from the simulation data without accounting for observational realism. We subdivide the profiles into strong cool-core, weak cool-core, and non-cool-core systems based on their central cooling time (see Section~\ref{sec:definition}). The three subgroups exhibit remarkable diversity in ICM thermodynamics, particularly for the gas within the central region ($<$\,100 kpc, see \cite{lehle.etal.2023}. Notably, there is a clear separation between SCCs and NCCs, with the former being characterised by higher core density, lower temperature, and lower entropy compared to the latter. In comparison with the XMM-Newton observed best-fit profiles of the Perseus cluster \citep{churazov.etal.2003}, the simulated SCC profiles largely resemble the observed ones: this is the case beyond the $\sim10$ kpc core in both normalisation and shape, and most remarkably for the entropy profiles. Within the core of 10 kpc, it is worth noting that the analytically approximated profiles (red thick curves) stop following the actual data of the gas density and temperature (red data points), which is  presumably because of the contamination of bright compact sources in the nucleus of NGC 1275 \citep{churazov.etal.2003}. As our X-ray modelling does not include such contaminated emission, the most robust comparison is between the simulated profiles and both the observed data and analytical fits beyond the innermost regions of Perseus. Overall, this comparison emphasises that the predicted ICM properties of the Perseus-like sample agree reasonably well with the X-ray observations of the Perseus cluster.  

\begin{figure*}
    \centering
    \includegraphics[width=0.9\textwidth]{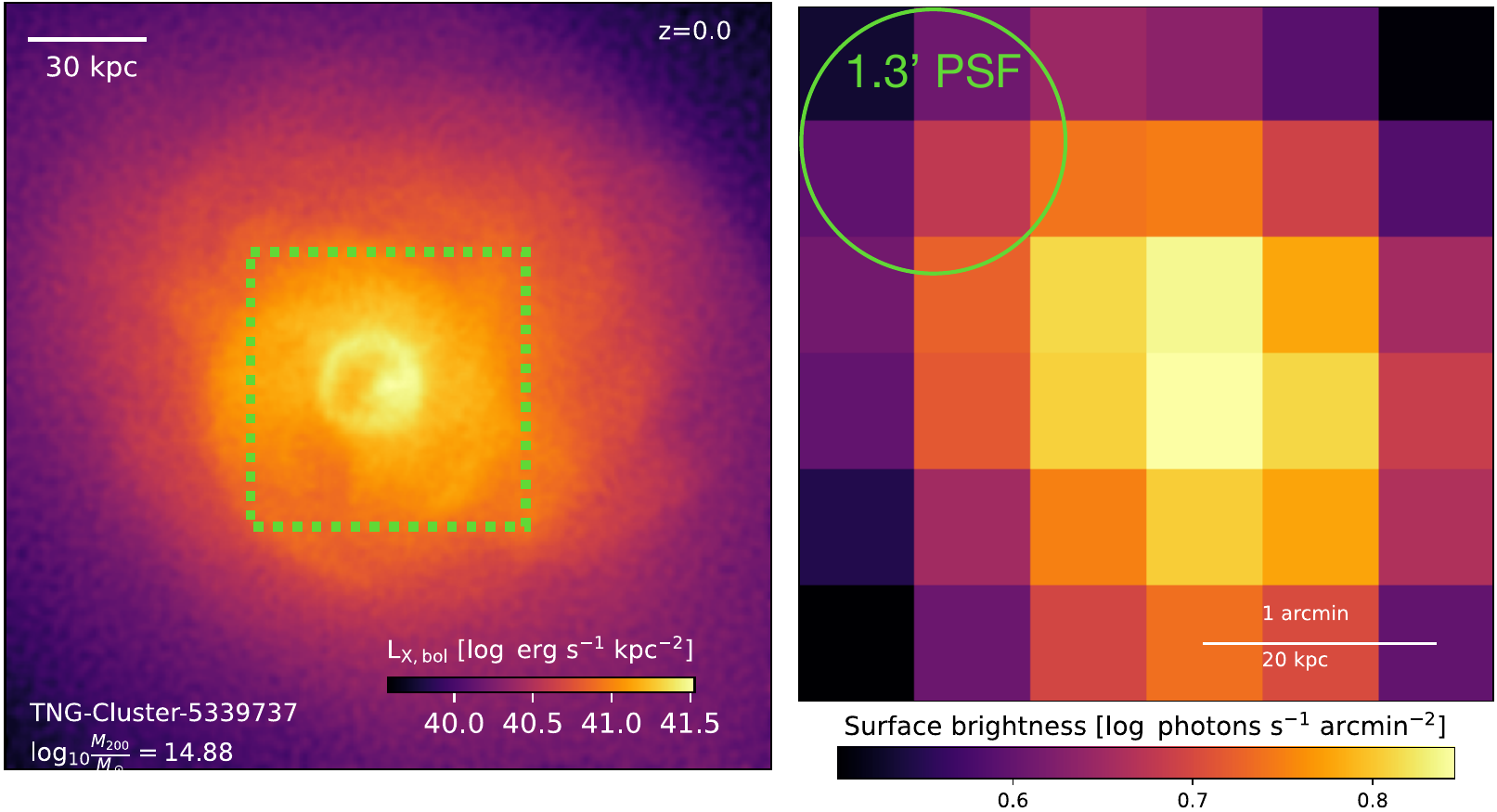}
   \includegraphics[width=0.99\textwidth]{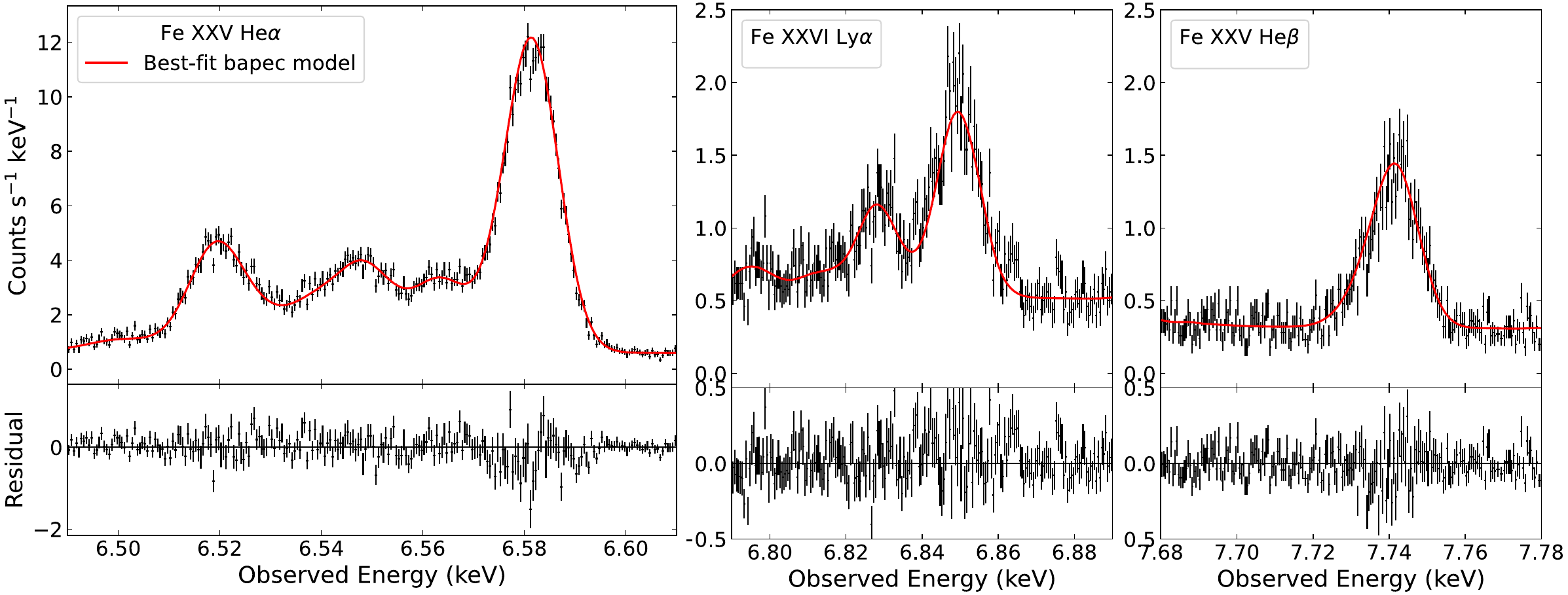}
    \caption{{\bf Illustration of the end-to-end mock X-ray analysis applied to TNG-Cluster to mimic observations with XRISM Resolve}. We showcase the process described in Section~\ref{sec:methods} for a single randomly selected Perseus-like cluster for an exposure of 100 ks. In the {\it top left} panel, we first show the intrinsic X-ray surface brightness map of the simulated cluster, which is similar to those of Fig.~\ref{fig:selection}, in which the central green square specifies the approximate FoV of XRISM Resolve ($\sim$ 70 x 70 kpc). The {\it top right} panel then shows the count-rate map via the mock observation. {\it Bottom:} Segments of the associated mock spectrum that resolve individual lines at the Fe XXV He$\alpha$ complex, Fe XXVI Ly$\alpha$, and Fe XXV He$\beta$, with the best-fit bapec model overlaid.}
    \label{fig:xray_mock}
\end{figure*}
\begin{figure*}
    \centering
    \includegraphics[width=0.79\textwidth]{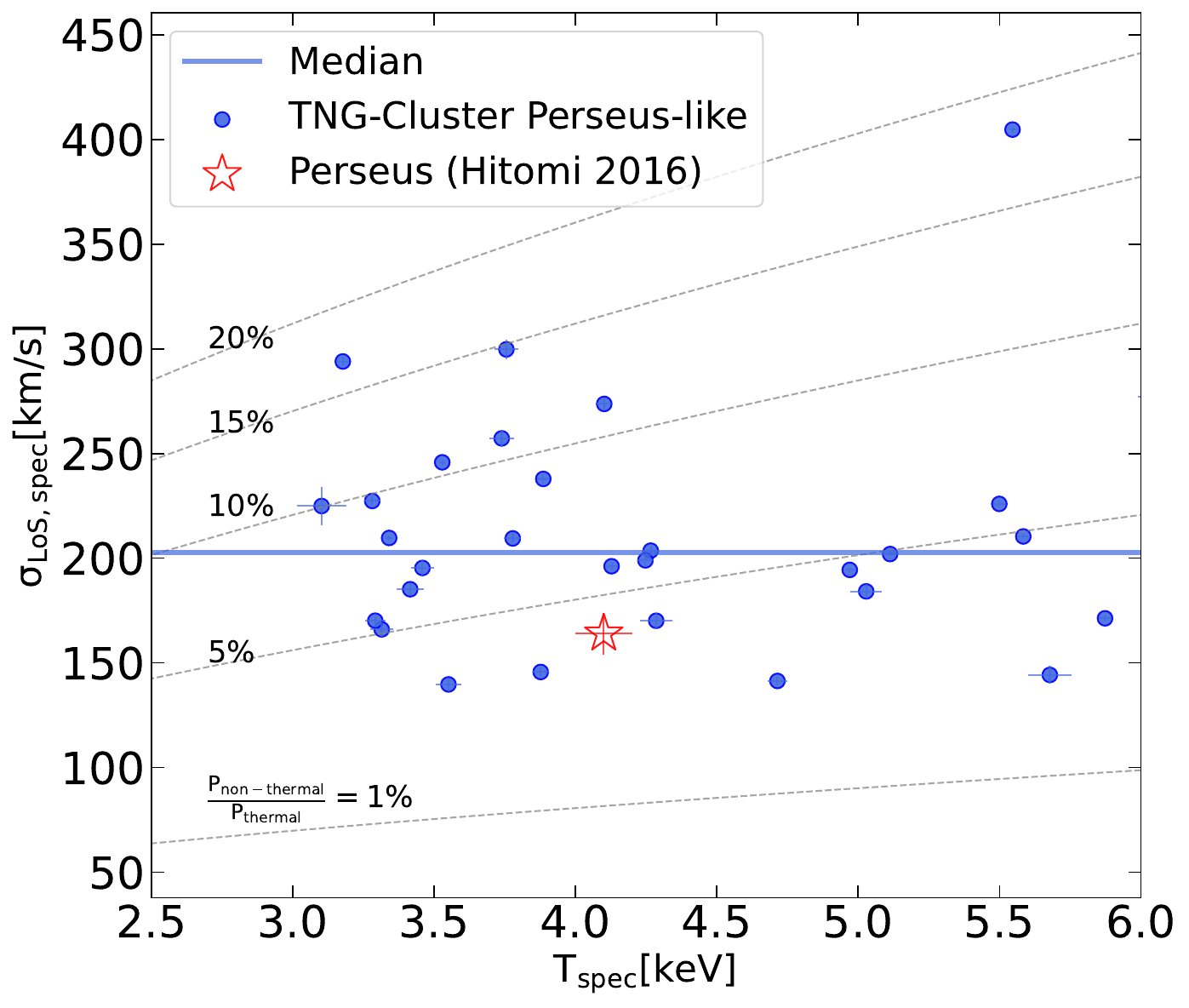}
    \caption{{\bf Relation predicted by TNG-Cluster and obtained via forward-modelling between best-fit velocity dispersion and best-fit temperature of the ICM in the central regions of Perseus-like clusters} in comparison to the Hitomi observation of Perseus \citep{hitomi.collaboration.2016}. These values are obtained by fitting mock spectra in the [6.4-8] keV band of 30 simulated cluster cores seen in a random orientation with 100 ks exposures with XRISM Resolve. The blue solid line gives the median value across the Perseus-like sample. In the background, we indicate with dashed lines the ratio ${\rm P_{non-thermal}/P_{thermal}}$, which is the ratio of motion-induced non-thermal pressure to thermal pressure (see text for details). TNG-Cluster naturally returns values for the LoS velocity dispersion in Perseus-like cores that are consistent with those derived for Perseus, with a low contribution of gas motions to the non-thermal compared to thermal pressure.}
    \label{fig:sigma_vs_kt}
\end{figure*}

In the left column of Fig. \ref{fig:xray_maps}, we showcase the X-ray surface brightness (SB) maps of a representative subset of Perseus-like clusters from TNG-Cluster. The SB maps show the X-ray bolometric emission ([0.1-100] keV) emitted by gas within the cluster core ($<$\,100 kpc). To enhance the visualisation of X-ray morphological features, we apply a Gaussian gradient magnitude (GGM) filter to the SB maps to produce SB gradient maps on a particular spatial scale\footnote{For the GGM filter, we employ the implementation from the scipy package (http://scipy.org/) with a smoothing length of ${\rm \sigma=2}$ pixels (2 kpc). The resulting gradient maps are displayed in the right column. The GGM method is recognised for its effectiveness in detecting edges and discontinuities in the X-ray surface brightness of galaxy clusters (\citealt{sanders.etal.2016}).}. Interestingly, the majority of these Perseus analogues display highly disturbed core regions characterised by various morphological features, including ripples, shock fronts, and cavities/bubbles (i.e. under-luminous regions). Some features in some of these simulated clusters are morphologically similar to those observed in Perseus \citep[e.g.][]{fabian.2012}. Whether or not all these manifestations of the simulated ICM are realistic remains to be determined. However, within the TNG model, it is clear that at least some of these morphological features are driven by the feedback released by the central SMBH. For example, Milky Way and Andromeda-like galaxies from TNG50 exhibit eROSITA-like features driven by SMBH feedback, with prominent shocks at the edges of over-pressurised and supervirial hot gas \citep{pillepich.etal.2021, truong.etal.2023}. We defer a more thorough and quantitative analysis of these X-ray SB fluctuations in Perseus-like clusters ---and in TNG-Cluster in general--- to future studies. 

Overall, the selected Perseus-like clusters from TNG-Cluster reproduce both the thermodynamics and the X-ray morphologies observed in the Perseus cluster reasonably
well, at least qualitatively. It is worth emphasising that, in this work, we do not aim to quantitatively compare the temperature, density, and entropy profiles, as well as the morphological features in resolved X-ray images, of the simulated clusters to Perseus. This is beyond the scope of the present work, which focuses on kinematics. In particular, here we present a detailed X-ray forward-modelling for XRISM, which we use to understand the observational inferences of kinematics. A similarly detailed mock X-ray study tailored to previous Chandra and XMM-Newton observations would be required to compare profiles and maps between TNG-Cluster and Perseus, and other observed systems: this a topic for future work.

\subsection{X-ray-based velocity dispersion of Perseus-like clusters}
\label{sec:Xray_measurements}

Given the qualitative consistency between the Perseus-like clusters simulated with TNG-Cluster and the real Perseus, we proceed by measuring the ICM kinematics of the former for comparison and interpretation of the latter. 

In Fig.~\ref{fig:xray_mock} we illustrate the end-to-end X-ray analysis of a Perseus-like cluster from the simulation, as described in Section~\ref{sec:methods}. The XRISM mock observation is produced assuming that the simulated cluster is located at $z=0.018$, which is the Perseus redshift, and with an exposure time of 100 ks. The images are centred on the cluster centre, defined as the point of minimum gravitational potential. Given the XRISM field of view (FoV) of about 3'x3', at $z=0.018$ XRISM probes a physical region within 70 kpc x 70 kpc in the cluster core, as shown by the green square in the top-left panel. In the top-right panel, we show the pixel-by-pixel count-rate map convolved with the XRISM point source function (PSF), which is $\sim1.3$'. The low spatial resolution of  XRISM makes it difficult to resolve X-ray morphological features that are clearly present in the intrinsic surface brightness map on the left.

Next we extract the spectrum from the mock observation and fit it with a bapec model to estimate the best-fit X-ray quantities. We consider all photons from the entire FoV excluding those within the core of 10 kpc, and postpone a spatially resolved analysis to future papers. The core exclusion aligns with observational practices, where the core is excluded to avoid contamination from the central AGN (e.g. \citealt{churazov.etal.2003}). Another reason for this choice is related to the fact that, within our analysis, we do not model the emission of the central SMBH. We checked that our results on the LoS velocity dispersions are not significantly affected by including or not the 10 kpc innermost regions. As shown in the bottom panel, for the fit we use a segment of the mock spectrum in the [6.4-8.0] keV band that covers the most prominent lines, including the FeXXV He$\alpha$ complex and, at higher energies, the FeXXVI Ly$\alpha$ and FeXXV H$\beta$ lines.

As we can see in the case of this particular cluster core, the bapec model is a good fit to the mock spectra. The reduced chi-squared value is $\sim 1$, suggesting that the profiles of the FeXXV He$\alpha$ complex and the other considered lines are approximately described by a single-temperature emission model. We verified that this is also the case in all other $30$ Perseus-like clusters.

We fitted the forward-modelled spectra in the [6.4-8.0] keV band of all selected TNG-Cluster systems. Figure~\ref{fig:sigma_vs_kt} shows the best-fit values of the LoS velocity dispersion and temperature of the ICM in the central regions of our Perseus-like cluster sample.

The derived LoS velocity dispersion represents, for each simulated cluster, the typical turbulent motions of the majority of its gas mass in the selected regions. This varies between 140 km/s and 300 km/s, with one cluster exhibiting a 400 km/s velocity dispersion in its core. Across the selected mass (i.e. temperature range), the median value of the velocity dispersion across the entire Perseus-like sample is $\sim 200$\,km/s. We  find, although we do not show, that the average velocity dispersion (emission-weighted) across all TNG-Cluster systems ---that is, not only Perseus-like ones--- is only slightly larger, and in all cases is $\lesssim 300$ km/s. This result aligns with the core velocity dispersion estimates for all the TNG-Cluster systems reported in the accompanying study by \cite{ayromlou.etal.2023}. Their study also highlights a substantial diversity in gas velocity dispersion across the entire TNG-Cluster sample. 

The Hitomi observation of Perseus \citep[][shown by the red star]{hitomi.collaboration.2016} falls well within the ballpark of the outcome of TNG-Cluster in this ${\rm \sigma_{LoS,spec}-T_{\rm spec}}$ diagram. We note that the Hitomi observation does not exactly probe the central core region of Perseus, as we consider in our analysis. Instead, it appears to have targeted an off-centre region northwest of the Perseus core. 

Based on the measured velocity dispersion, we can estimate the motion-induced pressure, which is provided by
\begin{equation}
    P_{\rm non-thermal}= \frac{1}{3}\rho\sigma^2, \label{eqn:p_nth}
\end{equation}
where $\rho$ is the gas mass density and $\sigma$ is the total gas velocity dispersion. Assuming that the gas motion is isotropic, $\sigma^2=3\sigma_{\rm LoS}^2$, we can compare this value to the thermal pressure:
\begin{equation}
    P_{\rm thermal}= nk_{\rm B} T, \label{eqn:p_th}
\end{equation}
where $n$ and $T$ are the gas number density and temperature, respectively. From equation (\ref{eqn:p_nth}) and (\ref{eqn:p_th}), one can deduce
\begin{equation}
    P_{\rm non-thermal}/P_{\rm thermal}=\frac{\rho \sigma_{\rm LoS}^2}{nk_B T} = \frac{\mu m_p\sigma_{\rm LoS}^2}{k_{\rm B} T}, \label{eqn:pratio}
\end{equation}
where $\mu$ is the mean molecular weight and $m_p$ is the proton mass. For our simulated Perseus-like sample, the contribution of the non-thermal pressure induced by gas motion is moderate compared to the thermal contribution. The pressure ratio is less than $15\%$ in more than $90\%$ of our Perseus-like clusters, and one-third of these clusters have a ratio of below $5\%$.  

\section{Interpretation of the ICM velocity dispersions}
\label{sec:intepretation}

Overall, TNG-Cluster supports the picture of subsonic motions in the core of the Perseus cluster as observed by Hitomi, whereby the motion-induced turbulent pressure is at the level of only 5-15 percent of the thermal pressure (Fig.~\ref{fig:sigma_vs_kt}). This implies that Perseus-like systems in TNG-Cluster have an ICM in the central regions that is largely in hydrostatic equilibrium. Nevertheless, the feature-rich and complex X-ray maps of the same simulated and observed cluster cores (Fig.~\ref{fig:xray_maps}) suggest a non-negligible level of activity from the central SMBHs. We are therefore interested in what values of 150-200 km/s for the ICM velocity dispersion tell us about the overall gas kinematics and the activity of the central engines. We quantify and discuss the answer below.
 
\subsection{Systematic uncertainties of the X-ray measurements}
\label{sec:systematics}

The first question to address is whether or not spectral fitting-derived values of velocity dispersion reflect intrinsic quantities, and if so, how accurately. We estimate the intrinsic velocities of the TNG-Cluster cores as emission-weighted or gas-mass-weighted average LoS velocity dispersions (Section~\ref{sec:definition}). We compute both using the gas cells in the central region with a side of 70 kpc, which is the approximate FoV of XRISM, as in Figs.~\ref{fig:xray_mock} and \ref{fig:sigma_vs_kt}, while excluding the innermost region with a radius of 10 kpc for consistency with the spectral-fitting measurements. The emission-weighted estimate takes emission in the narrow band of [6.0-8.0] keV, which covers all the prominent lines ---including the FeXXV He$\alpha$ complex--- that are used to derive the spectral-fitting ${\rm \sigma_{LoS}}$. It is noted that the energy range of the intrinsic emission extends slightly beyond the range used for spectral fitting ([6.4-8.0] keV). Nonetheless, we verify that this minor discrepancy does not substantially affect the results presented in this paper.

\begin{figure}
    \centering
    \includegraphics[width=0.47\textwidth]{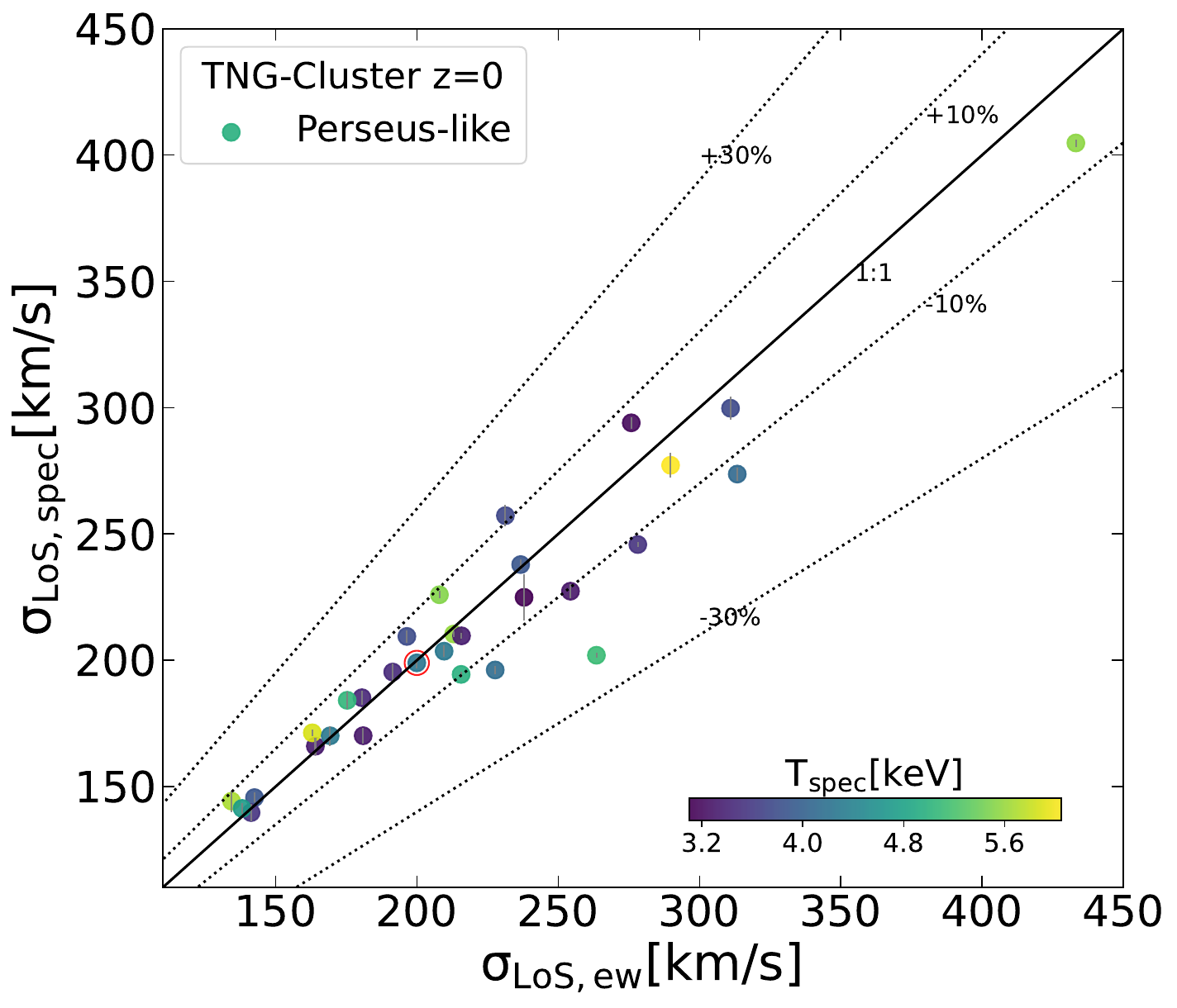}
    \includegraphics[width=0.47\textwidth]{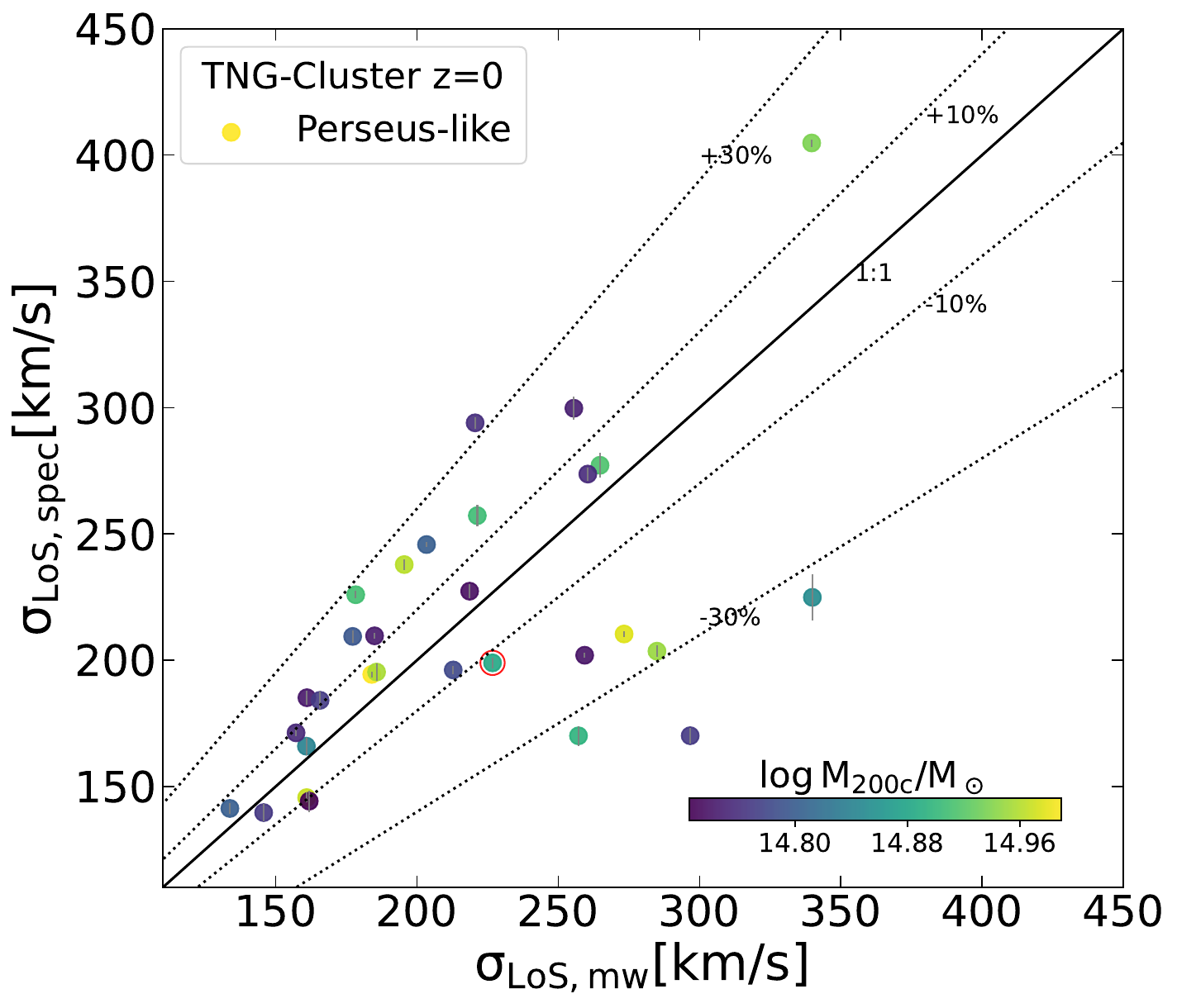}
    \caption{{\bf Intrinsic versus X-ray inferred velocity dispersion of the gas in cluster cores.} We compare the LoS spectral-fit values with intrinsic, emission-weighted ({\it top}) and mass-weighted ({\it bottom}) LoS velocity dispersion. The data points are colour coded according to either ${\rm T_{spec}}$ or ${\rm M_{200c}}$. The intrinsic values use all gas within a core region with a side of 70 kpc ---the XRISM field of view--- but excluding the innermost volume with radius of 10 kpc. The solid lines specify the 1:1 relation, whereas the dotted lines present the $\pm10\%$ and $\pm30\%$ deviations from the 1:1 relation. The red circle denotes the example cluster whose spectrum is shown in Fig.~\ref{fig:xray_mock}. }
    \label{fig:mock_vs_true}
\end{figure}

\begin{figure*}
    \centering
    \includegraphics[width=0.49\textwidth]{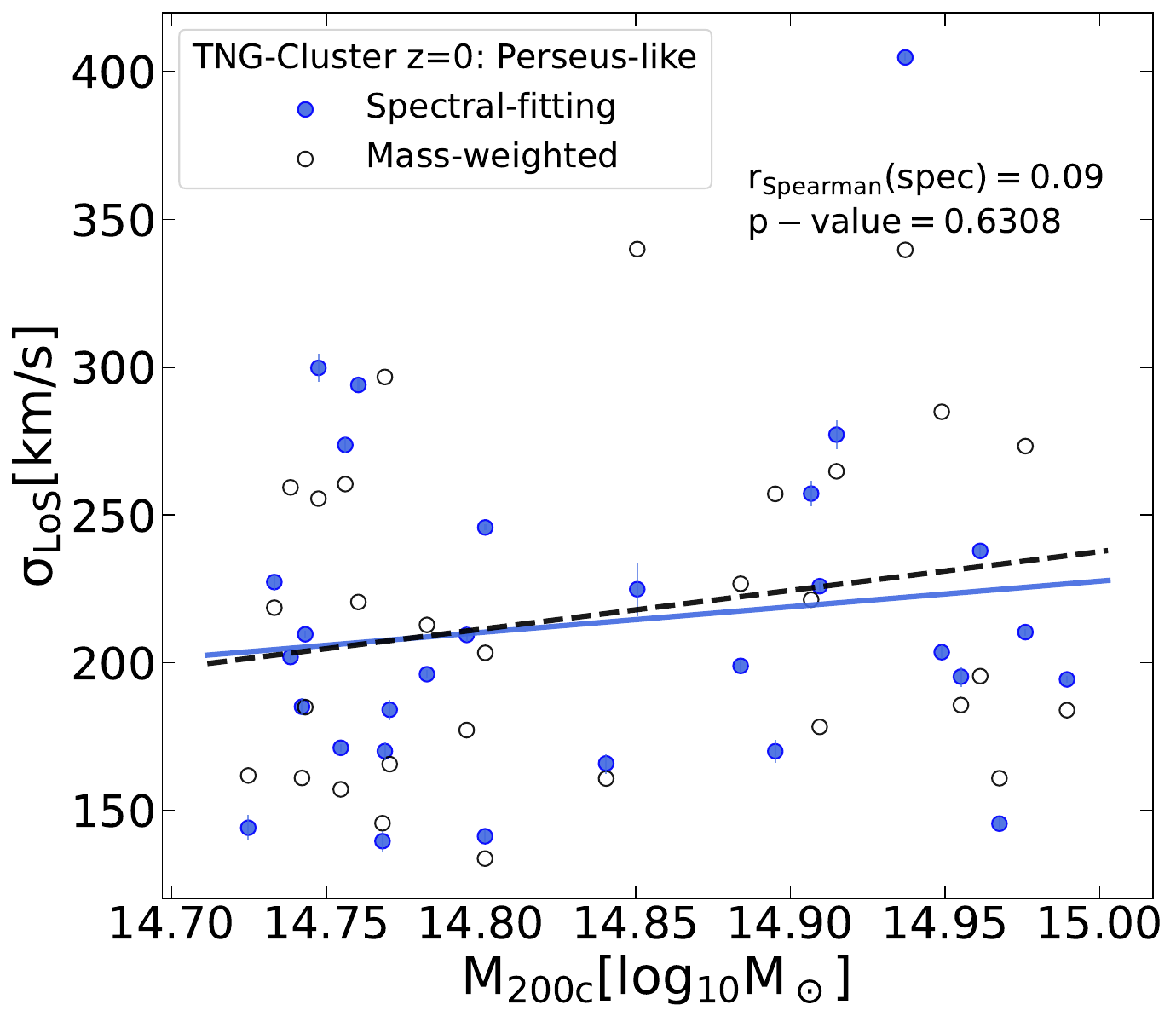}
    \includegraphics[width=0.49\textwidth]{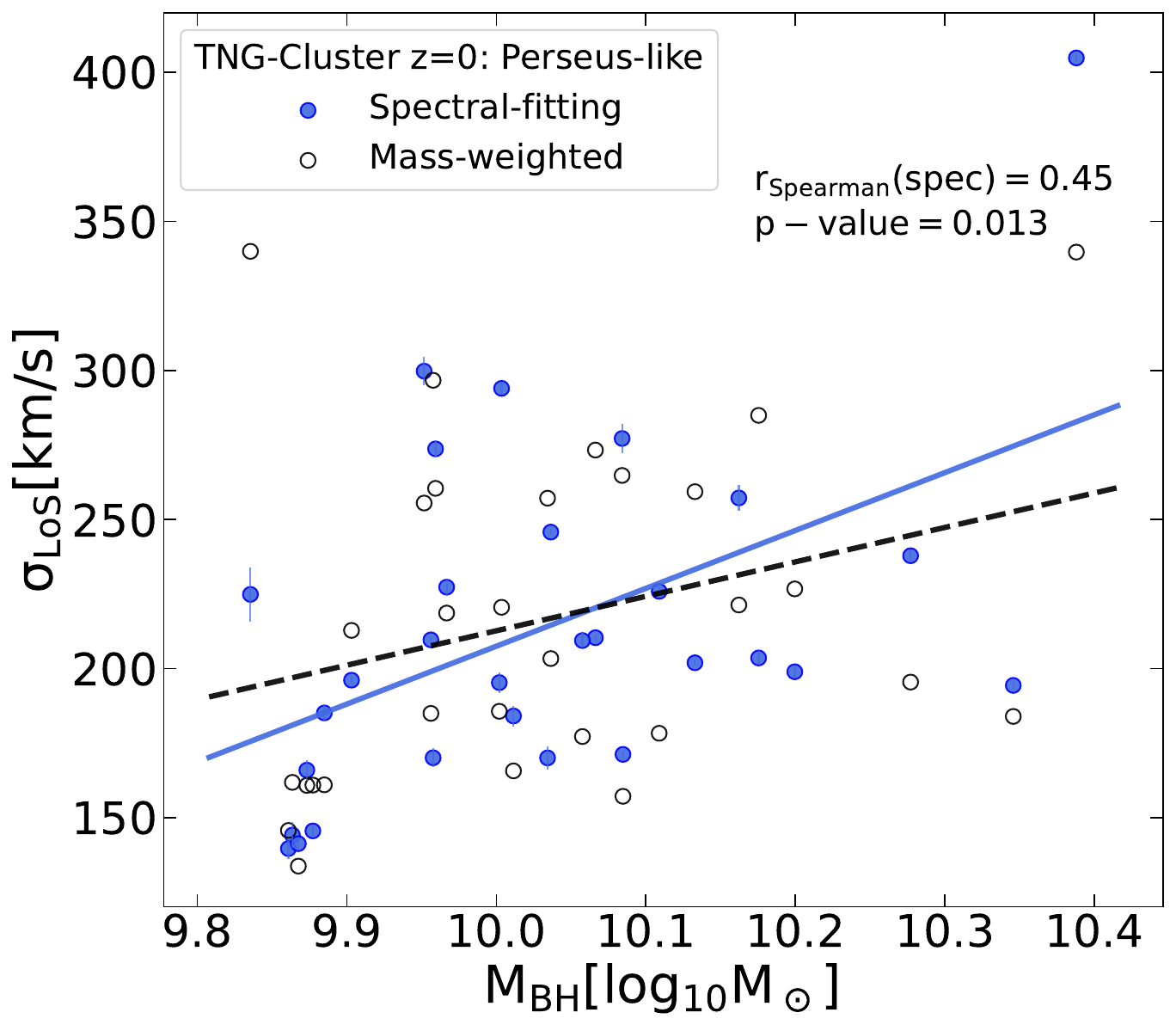}
    \includegraphics[width=0.49\textwidth]{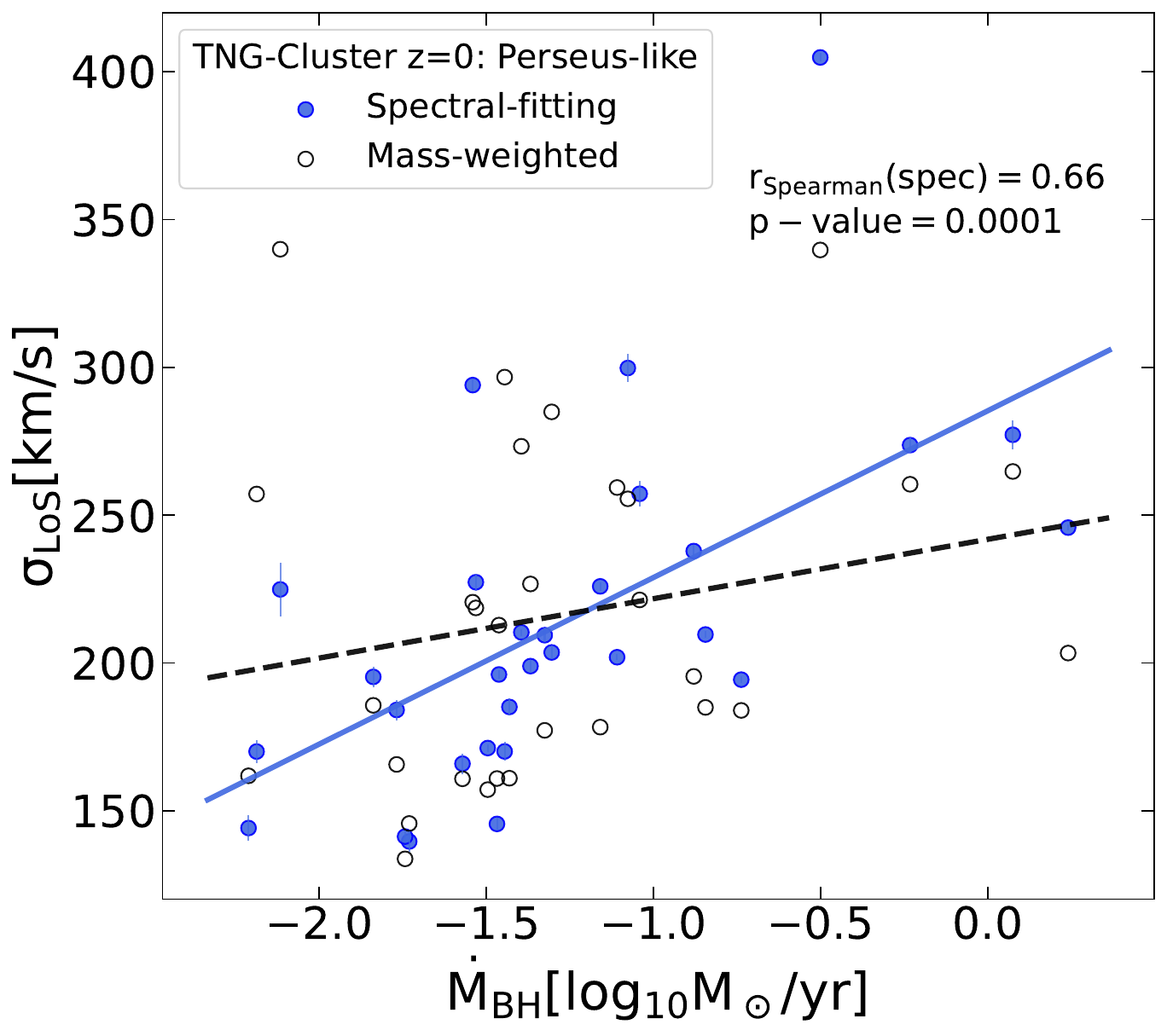}
    \includegraphics[width=0.49\textwidth]{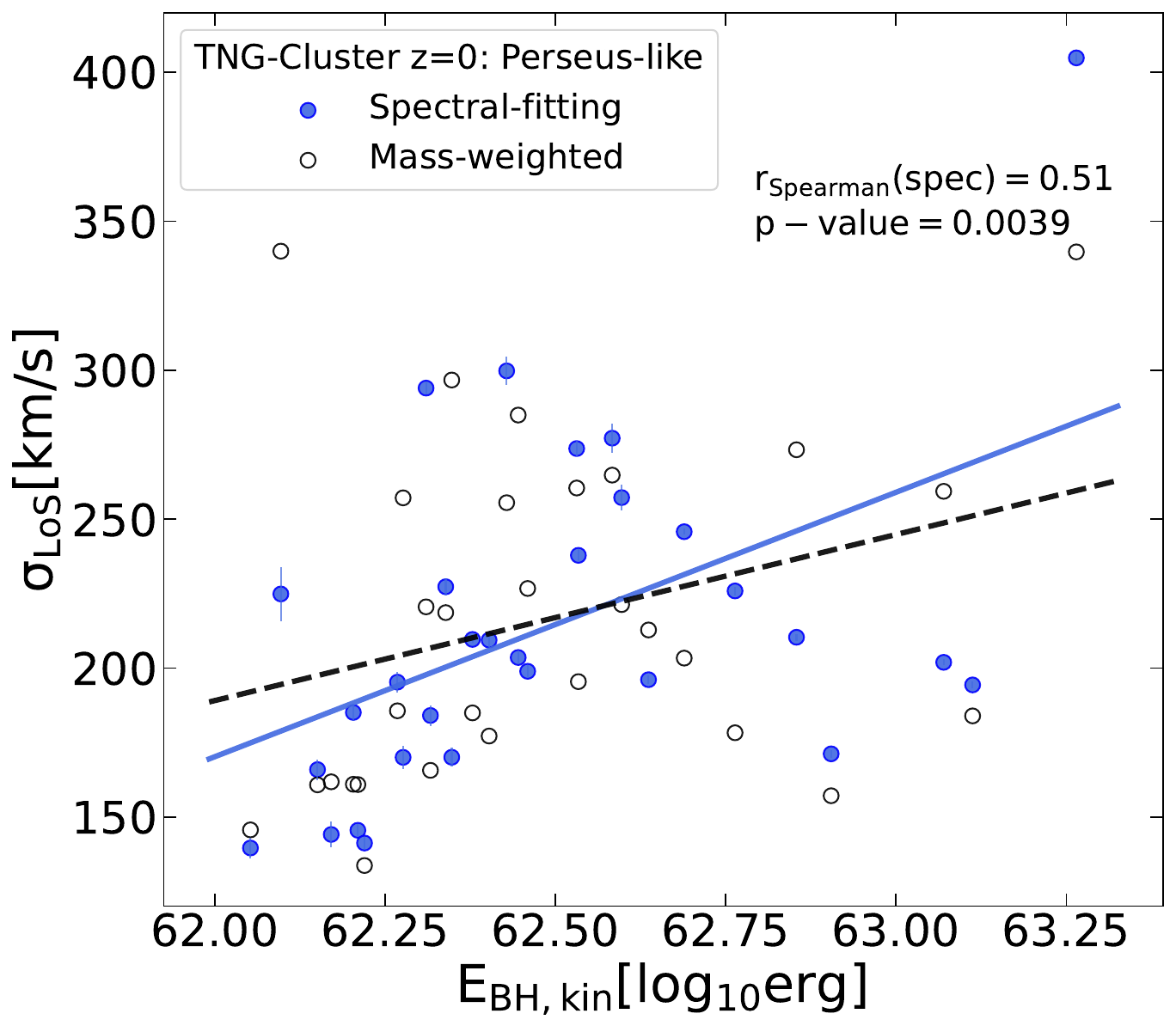}
    \caption{{\bf Dependence of the X-ray-inferred velocity dispersion of the ICM on cluster mass ($M_{200c}$) and central SMBH properties according to TNG-Cluster}. For the latter, we show SMBH mass ($M_{\rm BH}$, the single largest in each halo), instantaneous accretion rate ($\dot{M}_{\rm BH}$), and accumulated kinetic feedback energy ($E_{\rm kin}$). The solid (dashed) line represents the best-fit linear relation between the spectral-fitting (mass-weighted) velocity dispersion and cluster mass or SMBH properties. Cool-core clusters in the Perseus mass range exhibit larger levels of turbulence if they host a more massive SMBH, a SMBH that is accreting at higher rates, or a SMBH that has released more kinetic feedback on its surroundings throughout its life time.}
    \label{fig:dependence_m200_BH}
\end{figure*}

As shown in the top panel of Fig.~\ref{fig:mock_vs_true}, the values of the spectral-fit velocity dispersion correlate well with the emission-weighted estimates for all but one of the simulated Perseus-like clusters; that is, they agree to within $\pm10\%$ (dotted lines). Conversely, the spectral-fitting estimate may underestimate the true, mass-weighted intrinsic motions of the core gas by up to 100 km/s for systems where $\sigma_{\rm los,mw}$ is greater than 250 km/s. The discrepancy does not exhibit a clear correlation with halo mass (${\rm M_{200c}}$). The mismatch between ${\rm \sigma_{LoS, spec}}$ and ${\rm \sigma_{LoS,mw}}$ may originate from the inherent multi-phase nature of the ICM within these systems: relying exclusively on narrow spectral bands centred around a few prominent lines, such as the FeXXV He$\alpha$ complex, might not be sufficient to fully capture all gas with significant motion.
Regardless, the observationally derived gas velocity dispersion captures the typical true level of turbulence (i.e. the gas motion in cluster cores) to better than 30 per cent for the majority of halos.


\subsection{Dependence on cluster mass and SMBH properties}
\label{sec:m200_mbh_dependence}

We proceed to explore how the level of turbulence depends on cluster properties, particularly the activity of the central SMBHs. Within the TNG model, we expect the SMBHs at the centres of massive haloes to be active and to drive high-velocity outflows \citep{weinberger.etal.2017, weinberger.etal.2018, nelson.etal.2019b}. Although the majority of such SMBHs have low accretion rates, they occasionally inject significant amounts of kinetic energy, offsetting cooling, and we also expect this to be the case in TNG-Cluster halos (see also \citealt{ayromlou.etal.2023}).

In Fig.~\ref{fig:dependence_m200_BH}, we therefore show the relationship between spectral-fit, mass-weighted $\sigma_{\rm LoS}$ and cluster halo mass ($M_{200c}$), SMBH mass ($M_{\rm BH}$), SMBH instantaneous accretion rate ($\dot{M}_{\rm BH}$), and the kinetic feedback energy released by the SMBH throughout its lifetime (${\rm E_{kin}}$). With respect to halo mass, there is no substantial trend between $\sigma_{\rm LoS}$ and $M_{200c}$ within this narrow mass bin, as indicated by a Spearman correlation coefficient of ${\rm r_{Spearman}\approx0}$. However, we note that the {intrinsic} mass-weighted velocity dispersion of the gas mildly increases for more massive systems, shifting the sample-wide median from below 200 to 250 km/s across the Perseus-mass range.

\begin{figure*}
    \centering
    \includegraphics[width=0.99\textwidth]{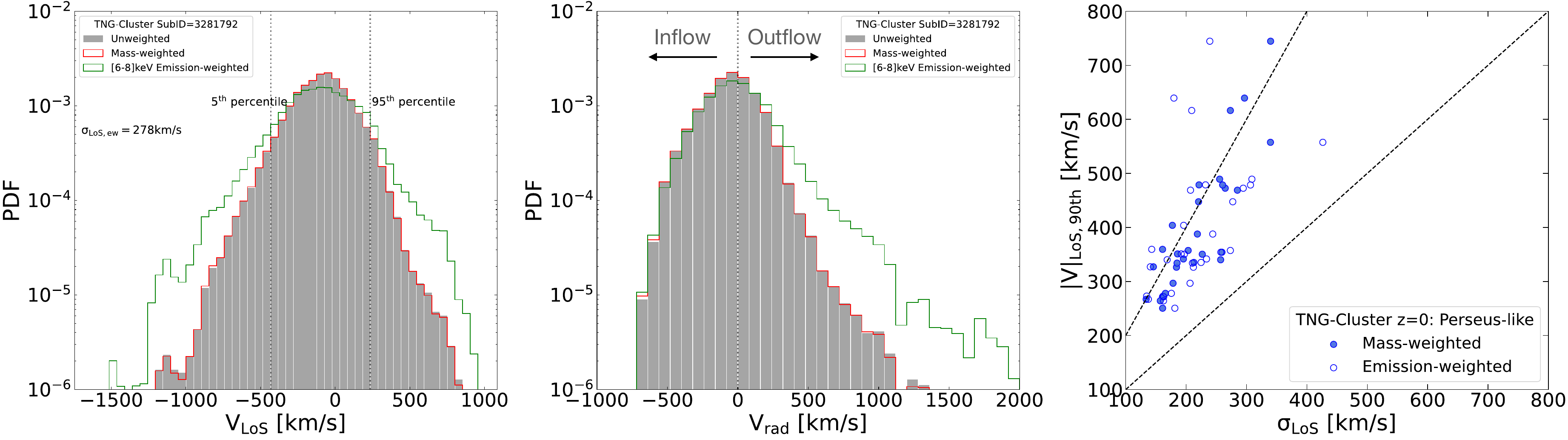}
    \includegraphics[width=0.99\textwidth]{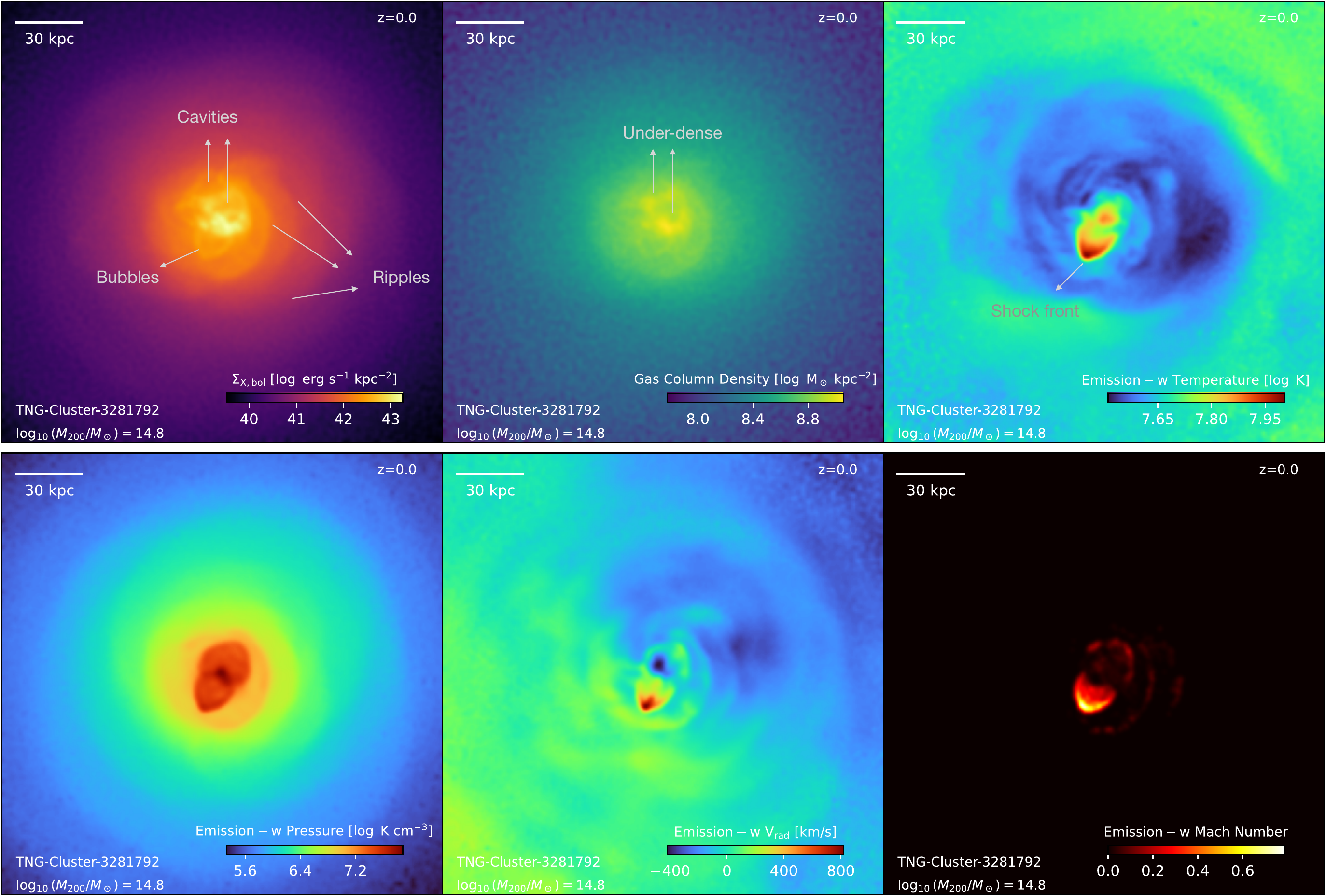}
    \caption{{\bf Connecting the line-of-sight velocity dispersion of the ICM in cluster cores to the velocity distribution, high-velocity outflows, and morphological features with TNG-Cluster.} In this figure, the top left/middle and bottom panels show one individual Perseus-like cluster, whereas the top right panel provides a statistical view across the entire TNG-Cluster Perseus-like sample at $z=0$. In particular, the {\it top left (middle)} panels show the 1D histogram of LoS (radial) velocity of the ICM within the core of 70 kpc. It demonstrates that bulk motions of relatively small fractions of the gas can make such distributions deviate from a Gaussian, such that the X-ray derived velocity dispersion does not fully capture the complexity of the velocity structure of the cluster. The {\it bottom} panel provides maps of the gas thermodynamical and kinematic properties (200 kpc a size for a full projection). Finally, the {\it top right} panel shows the correlation between the LoS velocity dispersion versus the 90th percentile of the absolute value of the LoS velocity, with the latter weighted by mass or X-ray emission. Dashed lines show the 1:1 and 2:1 relations. In TNG-Cluster Perseus-like systems, the velocity of the fastest-moving gas is greater than the velocity dispersion by approximately a factor of 2. Moreover, it is the fastest-moving gas that is chiefly responsible for the spatial features and fluctuations of the X-ray surface brightness maps.}
    \label{fig:vel_distribution_outflows}
\end{figure*}

Regarding the dependence on SMBH properties, a moderately positive correlation (${\rm r \gtrsim 0.5}$) is predicted by TNG-Cluster between the inferred LoS velocity dispersion and, particularly, SMBH accretion rate and the cumulative kinetic feedback energy. Except for a few halos, Perseus-like clusters with more massive SMBHs, SMBHs that are more highly accreting, and SMBHs that have released more cumulative kinetic feedback energy have larger levels of turbulent motions in their cores. These SMBH dependencies also manifest in the intrinsic mass-weighted ${\rm \sigma_{LoS}}$, albeit at a lower level of significance. We note that, in general, cool-core clusters, that is, Perseus-like halos in TNG-Cluster, host SMBHs that are more massive and so have  released more total feedback energy than weak cool-core and non-cool-core clusters in the same halo mass range (not shown). Despite the apparently low levels of gas velocity dispersion, these correlations suggest that SMBH feedback impacts the turbulence of the ICM within the cluster cores to a non-vanishing and possibly observable degree.

\subsection{Connecting to high-velocity bulk motions in the cluster cores}
\label{sec:outflows}

We are interested in whether or not it is  possible that {low} inferred values of gas velocity dispersion may coexist with  higher velocity bulk motions of a small fraction of the ICM mass. In other words, we would like to know how well the gas velocity dispersion describes the whole velocity distribution of the ICM in cluster cores. The interpretation of the measured velocity dispersion and what it can tell us about SMBH feedback, for example, is complicated by the multi-phase and multi-velocity nature of the gas in clusters, as we show below.

In Fig.~\ref{fig:vel_distribution_outflows}, we use another individual Perseus-like cluster to demonstrate the complexity of the ICM velocity field and how it connects to the X-ray surface brightness maps. In particular, the top left panel of Fig.~\ref{fig:vel_distribution_outflows} shows the distribution of the intrinsic LoS velocity (${V_{\rm LoS}}$) in the core region of a Perseus-like cluster, that is, both mass and emission weighted. The depicted cluster is characterised by an emission-weighted velocity dispersion of about 278 km/s. However, for this system and indeed for a significant fraction of our Perseus-like clusters, a Gaussian distribution does not adequately capture the $V_{\rm LoS}$ distribution, and in particular it underestimates its tails. To quantify the tail significance, we compute the kurtosis excess for the LoS velocity distribution and find that about $43\%$ of the Perseus-like clusters exhibit a significant excess kurtosis (>1). This indicates significant outliers in the tails beyond that expected for a Gaussian distribution. 

Upon inspecting the radial velocity (${\rm V_{rad}}$) distribution, as depicted in the middle panel of Fig.~\ref{fig:vel_distribution_outflows}, it is evident that it exhibits a pronounced skew toward positive values, indicative of outflowing gas. This is particularly apparent when the distribution is weighted by X-ray emission. This result suggests that outflowing X-ray-emitting gas, capable of reaching velocities of up to 1000 km/s, is primarily responsible for the extended tails of the LoS velocity distribution. This pattern is consistent across the majority of clusters in the Perseus-like subsample of TNG-Cluster. It is worth mentioning that, in some systems, we can also find  inflowing gas with velocities exceeding 1000 km/s (see more detailed analysis in \citealt{ayromlou.etal.2023}). Nonetheless, the X-ray emission in the [6-8] keV band of this gas is often negligible in comparison to the outflow emission, which is likely due to their relatively low temperatures. 

According to TNG-Cluster and across a large fraction of the Perseus-like subsample, it appears that the fast outflows predominantly shape the tails of the velocity distributions in the core regions, with velocities reaching several hundreds of km/s. This is in contrast to a mean LoS velocity of about 100 km/s and an inferred velocity dispersion of about 200 km/s on average across the Perseus-like clusters. Quantitatively, in the top right panel of Fig.~\ref{fig:vel_distribution_outflows} we show the 90th percentile of the $|V_{\rm LoS}|$ distribution in comparison with the $\sigma_{\rm LoS}$ (i.e.  approximately the width of the $V_{\rm LoS}$ distribution) for the Perseus-like subsample. Clearly, cluster cores that are more turbulent based on the value of their velocity dispersion also exhibit high-velocity tails. However, and notably, the velocity of fast-moving gas is greater than the velocity dispersion by a factor of $\gtrsim 2$. In other words, a small fraction of the gas mass in cluster cores moves much faster than the bulk of the ICM. 

In order to investigate the influence of these high-velocity components, in the bottom panels of Fig.~\ref{fig:vel_distribution_outflows} we trace the origin of the morphological features seen in the X-ray maps by comparing with maps of the thermodynamical and velocity fields of the same gas. In particular, we show projected maps of X-ray surface brightness, gas column density, temperature, gas pressure, radial velocity, and Mach number within a core region of 200 kpc on one side and with a full projection depth. 

The X-ray cavities and bubbles within the central region (<30 kpc) of this simulated cluster are associated with under-dense regions. Furthermore, the bubble borders are co-spatial with shock fronts, which are characterised by distinct temperature jumps with respect to the ambient medium. Beyond the core, X-ray ripples featuring spiral or dome-like shapes as seen in the brightness map appear to be created by hot and dense gas contained within thin shells, which  are clearly visible in the pressure map. Most importantly, these shells move outward faster than the rest of the ICM as evidenced by the radial velocity map. Notably, within the 30 kpc core, fast outflows reaching speeds above 500 km/s significantly increase the gas temperature to supervirial levels and generate shock fronts within the core.

We note that the cluster shown in Fig.~\ref{fig:vel_distribution_outflows} is just one among many in TNG-Cluster. While it serves as a good example to illustrate the multi-phase nature of gas velocities in Perseus-like clusters, this cluster most certainly does not represent a replica of Perseus. In particular, the gas bulk motion in the cluster core appear to be, at face value, remarkably higher than the observed value of the Perseus cluster ($\sim100$ km/s, \citealt{hitomi.collaboration.2018}). However, firstly, in our simulated Perseus-like clusters, whereas the outflows could have an emission-weighted radial velocity on average in the 300-700 km/s range, their relative contribution to the overall core emission is around 5-7$\%$, potentially rendering their detection via X-ray observations quite challenging. Secondly and most importantly, examining the LoS velocity map (not shown here) reveals that the average LoS velocity is approximately three to four times less than the radial velocity value. This is likely attributable to the anistropic nature of the outflows in combination with the projection effect. Finally, it is worth mentioning that there is notable cluster-to-cluster variation in gas bulk motions in the cores according to TNG-Cluster. For an accurate and fair comparison with observational data, a dedicated analysis would need to extend beyond face-value comparisons and would require a spatially resolved spectral analysis. As such an extensive comparison is beyond the scope of this paper, we defer this analysis to future works.

The picture suggested by the simulated cluster of Fig.~\ref{fig:vel_distribution_outflows}, and shared by most of the systems in our Perseus-like sample, is that the complex spatial fluctuations of the X-ray surface brightness are driven by the high-velocity tails of the ICM velocity distribution. These components can move as fast as many hundreds of km/s, even if the overall velocity dispersion, and hence turbulence of the ICM in the core, is only 100-200 km/s. 

Within the TNG model, such high-velocity outflows are driven by the energy injected by the central SMBHs in the kinetic feedback mode \citep{nelson.etal.2019b}. We demontrated, in different halos such as in Milky Way-like galaxies \citep{pillepich.etal.2021}), that the dynamical interaction between SMBH-driven outflows and halo gas can produce X-ray cavities, shells, and bubbles, the latter clearly co-spatial with shock fronts. We postpone the task of quantifying these connections in TNG-Cluster  to future work. However, we speculate that such SMBH-driven outflows sweep the ICM along their path, thereby creating bubbles or cavities; they also compress the ICM into shells, which are subsequently propelled outward initially at high velocities, ranging from several hundreds to thousands of km/s, creating shocks that provide a small-scale turbulent driving term. These shells gradually slow down over time, and their initial feedback-injected kinetic energy is partially converted into thermal energy, further modulating the temperature and cooling times of the ICM.

\begin{figure}
    \centering
    \includegraphics[width=0.45\textwidth]{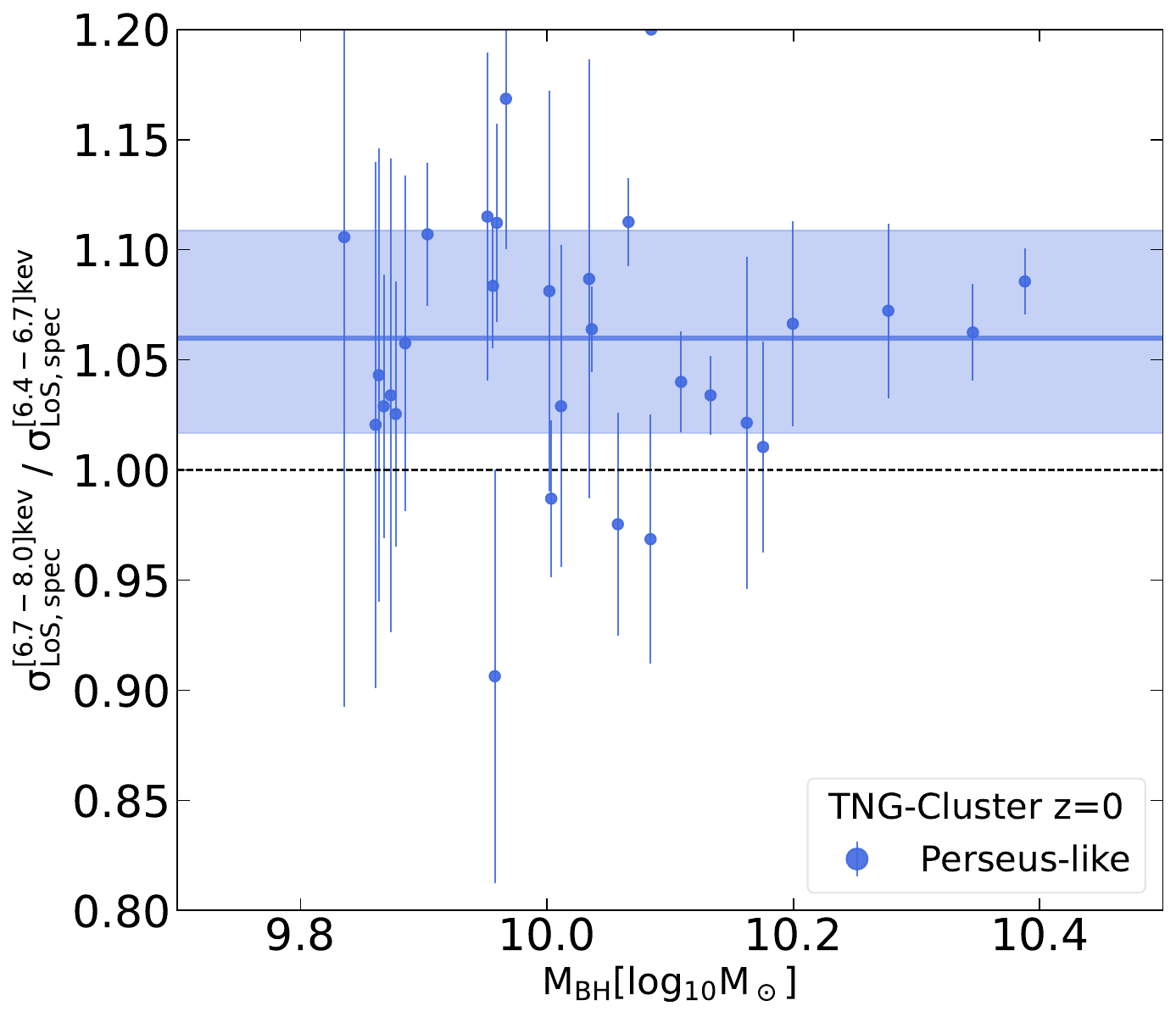}
    \caption{{\bf Comparison of the inferred velocity dispersion of the ICM in the core of clusters derived from low-energy vs. higher-energy spectral lines.} In particular, we plot the ratio of spectral-fit LoS velocity dispersions using two different X-ray bands: [6.7-8.0] keV versus [6.4-6.7] keV, as a function of SMBH mass. The solid line represents the median relation, whereas the shaded area specifies the ${\rm 16^{th}-84^{th}}$ envelope. For the Perseus-like sample, the ratio is systematically larger than 1, although by only 6-10 percent.}
    \label{fig:sigma_ratio}

\end{figure}

\subsection{Excess in the LoS velocity dispersion from high-energy lines}
\label{sec:sigma_ratio}

The high-velocity bulk motions in the ICM will amplify the inferred LoS velocity dispersion, either because of pure geometrical and projection effects, or because of these plus the physical enhancement of the ICM turbulence due to the shock fronts within the innermost regions. As we show above, this fast-moving gas has temperatures that can surpass the virial temperature of the clusters. This implies a possible bias; that is, an excess in the LoS velocity dispersion when inferred from high-energy X-ray lines.

To illustrate this point, we carry out an exercise wherein the spectral LoS velocity dispersion is measured by fitting mock spectra across two different energy bands, specifically [6.45-6.65] keV versus [6.7-8.0] keV in the observed frame of reference. We measure the line broadening using lines at distinct energies, and compare the measurements. Within the lower-energy band, the line-broadening measurement predominantly relies on the FeXXV He$\alpha$ complex, whose emission peaks around 6.70 keV (in the rest frame). For the higher-energy band, the broadening measurement chiefly relies on two lines: FeXXVI Ly$\alpha$ (6.98 keV, rest frame) and FeXXV He$\beta$ (7.88 keV, rest frame). 

Figure~\ref{fig:sigma_ratio} shows the ratio between the spectral-fit $\sigma_{\rm LoS}$ in the [6.7-8.0] keV range and that fit with the [6.45.6.65] keV range, as a function of SMBH mass for the TNG-Cluster Perseus-like systems. The ratio is systematically above unity, with a median excess in $\sigma_{\rm LoS}$ measured from higher-energy bands of about $6\%$. Notably, we find that there is a moderate correlation (${\rm r_{Spearman}\approx0.5}$) between the ${\rm \sigma_{LoS,spec}}$ ratio and the kurtosis of the LoS velocity distribution (not shown). This correlation is a manifestation of the multi-phase nature of the ICM. We find no significant correlation between the $\sigma_{\rm LoS}$ ratio and the SMBH mass. Although small, this effect reflects the multi-phase and complex kinematics and thermodynamical structure of the ICM in cool-core clusters.  

\section{Summary and conclusions}
\label{sec:summary}

In this paper, we present a study of the X-ray-derived kinematics of the intracluster medium (ICM) in the core of Perseus-like galaxy clusters simulated within the new TNG-Cluster simulation. In particular, we performed end-to-end XRISM mock observations of the simulated systems and fitted the X-ray spectra as in observations. We thereby determined the level of ICM turbulence in the core regions, that is, the LoS velocity dispersion, and assessed its dependence on the activity of the central SMBHs. Our main findings are summarised as follows:

\begin{enumerate}
    \item Given the abundant statistics and overall realism of the TNG-Cluster simulation suite, we identify 30 simulated clusters at $z=0$ that are similar to the Perseus cluster in both halo mass and cool-core nature. Moreover, the selected Perseus-like clusters have total X-ray luminosities, thermodynamical radial profiles, and X-ray surface-brightness maps (with bubbles, ripples, cavities, and shock fronts) that are quantitatively and/or qualitatively consistent with observations of Perseus (Figs.~\ref{fig:selection} and \ref{fig:xray_maps}).
    
    \item Across our Perseus-like sample, the median value of gas velocity dispersion measured from the mock X-ray analysis (Fig.~\ref{fig:xray_mock}) is about 200 km/s, with small cluster-to-cluster variation (140-300 km/s). This indicates that pressure from turbulent motion is less than 10$\%$ of the thermal pressure, on average. The Hitomi observations of velocity dispersion in Perseus fall within the predictions from TNG-Cluster (Fig.~\ref{fig:sigma_vs_kt}). 
    
    \item We confirm that the velocity dispersion derived via X-ray spectral fitting correctly captures (within $10\%$) the emission-weighted value. On the other hand, in certain cases, it underestimates the mass-weighted intrinsic velocity dispersion by up to $100$ km/s, albeit with a typical error of $< 30\%$ (Fig.~\ref{fig:mock_vs_true}).
    
    \item Within the narrow range of ICM velocity dispersion in their cores, Perseus-like clusters with more massive SMBHs, with SMBHs that are more highly accreting, and with SMBHs that have released more kinetic feedback energy all have larger levels of turbulent motion (Fig.~\ref{fig:dependence_m200_BH}). This suggests that SMBH feedback impacts ICM turbulence to some degree. 

    \item The low values of inferred ICM velocity dispersion coexist with high-velocity outflows and bulk motions of relatively small amounts of supervirial hot gas moving at hundreds to thousands of km/s. However, it should be noted that the detection of high-velocity outflows in observations may prove challenging, owing to their anisotropic nature in combination with projection effects. The coexistence of different velocity components is due to the multi-phase and multi-structure nature of the ICM, and to the complexity of the gas velocity distributions, of which the velocity dispersion only offers a partial description (Fig.~\ref{fig:vel_distribution_outflows}).
    
    \item TNG-Cluster predicts that these hot  high-velocity outflows influence the morphology seen in the X-ray surface-brightness maps, as well as the kinematics of the core ICM. As outflows propagate, they compress gas into thin, dense, and shock-heated shells in the innermost regions of the cores, contributing to the formation of the cavities, bubbles, and shock fronts visible in X-ray imaging. The degree to which these outflow-induced morphological features quantitatively align with observations remains to be determined. At the same time, the outflows amplify gas motions in the hot phase; this systematic uncertainty elevates the LoS velocity dispersion if inferred from higher-energy X-ray emission lines (Fig. \ref{fig:sigma_ratio}). 
\end{enumerate}

Overall, our findings with TNG-Cluster support the scenario of subsonic gas motions within the core of the Perseus cluster, as observed by Hitomi, with low levels of turbulent pressure compared to thermal pressure. However, relatively low values of inferred velocity dispersion do not imply dormant central SMBHs. Indeed, SMBHs can still drive bulk motions and high-velocity outflow, which in turn generate ICM turbulence, increasing the overall velocity dispersion. In addition, SMBH feedback is also responsible for producing many of the rich features of observed and simulated X-ray surface brightness maps. In future work, we will analyse the spatial distribution of 
ICM bulk motions within and beyond the clusters core, exploring the morphological and spatial complexity of the selected Perseus-like clusters in TNG-Cluster. The simulated results will be thoroughly compared with existing Hitomi observations as well as with upcoming observations from XRISM and future X-ray IFU instruments such as LEM, HUBS, and ATHENA.

\section*{Acknowledgements}

We would like to thank the anonymous Referee for a thorough review that helps improve the manuscript. The authors thank John ZuHone and Brian MacNamara for useful discussions and Eugene Churazov for kindly providing us with XMM-Newton observed data of the Perseus cluster. NT thanks Gerrit Schellenberger for helpful discussions. DN and MA acknowledge funding from the Deutsche Forschungsgemeinschaft (DFG) through an Emmy Noether Research Group (grant number NE 2441/1-1). KL acknowledges funding from the Hector Fellow Academy through a Research Career Development Award. KL is a fellow of the International Max Planck Research School for Astronomy and Cosmic Physics at the University of Heidelberg (IMPRS-HD). Moreover, this work is co-funded by the European Union (ERC, COSMIC-KEY, 101087822, PI: Pillepich). Views and opinions expressed are however those of the author(s) only and do not necessarily reflect those of the European Union or the European Research Council. Neither the European Union nor the granting authority can be held responsible for them. The material is based upon work supported by NASA under award number 80GSFC21M0002.

The TNG-Cluster simulation suite has been executed on several machines: with compute time awarded under the TNG-Cluster project on the HoreKa supercomputer, funded by the Ministry of Science, Research and the Arts Baden-Württemberg and by the Federal Ministry of Education and Research; the bwForCluster Helix supercomputer, supported by the state of Baden-Württemberg through bwHPC and the German Research Foundation (DFG) through grant INST 35/1597-1 FUGG; the Vera cluster of the Max Planck Institute for Astronomy (MPIA), as well as the Cobra and Raven clusters, all three operated by the Max Planck Computational Data Facility (MPCDF); and the BinAC cluster, supported by the High Performance and Cloud Computing Group at the Zentrum für Datenverarbeitung of the University of Tübingen, the state of Baden-Württemberg through bwHPC and the German Research Foundation (DFG) through grant no INST 37/935-1 FUGG. 

The IllustrisTNG simulations are publicly available and accessible at \url{www.tng-project.org/data} \citep{nelson.etal.2019a}. The TNG-Cluster simulation will also be made public on this platform in the near future. Data directly related to this publication are available on request from the corresponding author.
\bibliographystyle{aa}
\bibliography{refs}

\begin{thebibliography}{73}
\expandafter\ifx\csname natexlab\endcsname\relax\def\natexlab#1{#1}\fi

\bibitem[{{Aguerri} {et~al.}(2020){Aguerri}, {Girardi}, {Agulli}, {Negri},
  {Dalla Vecchia}, \& {Dom{\'\i}nguez Palmero}}]{Aguerri.etal.2020}
{Aguerri}, J.~A.~L., {Girardi}, M., {Agulli}, I., {et~al.} 2020, \mnras, 494,
  1681

\bibitem[{{Allen} {et~al.}(2011){Allen}, {Evrard}, \&
  {Mantz}}]{allen.evrard.mantz.2011}
{Allen}, S.~W., {Evrard}, A.~E., \& {Mantz}, A.~B. 2011, \araa, 49, 409

\bibitem[{{Ayromlou} {et~al.}(2023){Ayromlou}, {Nelson}, {Pillepich}, {Rohr},
  {Truong}, {Li}, {Simionescu}, {Lehle}, \& {Lee}}]{ayromlou.etal.2023}
{Ayromlou}, M., {Nelson}, D., {Pillepich}, A., {et~al.} 2023, arXiv e-prints,
  arXiv:2311.06339

\bibitem[{{Biffi} {et~al.}(2013){Biffi}, {Dolag}, \&
  {B{\"o}hringer}}]{biffi.etal.2013}
{Biffi}, V., {Dolag}, K., \& {B{\"o}hringer}, H. 2013, \mnras, 428, 1395

\bibitem[{{Biffi} {et~al.}(2012){Biffi}, {Dolag}, {B{\"o}hringer}, \&
  {Lemson}}]{biffi.etal.2012}
{Biffi}, V., {Dolag}, K., {B{\"o}hringer}, H., \& {Lemson}, G. 2012, \mnras,
  420, 3545

\bibitem[{{B{\"o}hringer} \& {Werner}(2010)}]{bohringer.werner.2011}
{B{\"o}hringer}, H. \& {Werner}, N. 2010, \aapr, 18, 127

\bibitem[{{Bourne} \& {Sijacki}(2017)}]{bourne.sijacki.2017}
{Bourne}, M.~A. \& {Sijacki}, D. 2017, \mnras, 472, 4707

\bibitem[{{Churazov} {et~al.}(2000){Churazov}, {Forman}, {Jones}, \&
  {B{\"o}hringer}}]{churazov.etal.2000}
{Churazov}, E., {Forman}, W., {Jones}, C., \& {B{\"o}hringer}, H. 2000, \aap,
  356, 788

\bibitem[{{Churazov} {et~al.}(2003){Churazov}, {Forman}, {Jones}, \&
  {B{\"o}hringer}}]{churazov.etal.2003}
{Churazov}, E., {Forman}, W., {Jones}, C., \& {B{\"o}hringer}, H. 2003, \apj,
  590, 225

\bibitem[{{Davis} {et~al.}(1985){Davis}, {Efstathiou}, {Frenk}, \&
  {White}}]{davis.etal.1985}
{Davis}, M., {Efstathiou}, G., {Frenk}, C.~S., \& {White}, S.~D.~M. 1985, \apj,
  292, 371

\bibitem[{{Donahue} \& {Voit}(2022)}]{donahue.voit.2022}
{Donahue}, M. \& {Voit}, G.~M. 2022, \physrep, 973, 1

\bibitem[{{Donnari} {et~al.}(2021){Donnari}, {Pillepich}, {Nelson},
  {Marinacci}, {Vogelsberger}, \& {Hernquist}}]{donnari.etal.2021b}
{Donnari}, M., {Pillepich}, A., {Nelson}, D., {et~al.} 2021, \mnras, 506, 4760

\bibitem[{{Fabian}(2002)}]{fabian.2002}
{Fabian}, A.~C. 2002, in Lighthouses of the Universe: The Most Luminous
  Celestial Objects and Their Use for Cosmology, ed. M.~{Gilfanov},
  R.~{Sunyeav}, \& E.~{Churazov}, 24

\bibitem[{{Fabian}(2012)}]{fabian.2012}
{Fabian}, A.~C. 2012, \araa, 50, 455

\bibitem[{{Fabian} {et~al.}(2011){Fabian}, {Sanders}, {Allen}, {Canning},
  {Churazov}, {Crawford}, {Forman}, {Gabany}, {Hlavacek-Larrondo}, {Johnstone},
  {Russell}, {Reynolds}, {Salom{\'e}}, {Taylor}, \& {Young}}]{fabian.etal.2011}
{Fabian}, A.~C., {Sanders}, J.~S., {Allen}, S.~W., {et~al.} 2011, \mnras, 418,
  2154

\bibitem[{{Giacintucci} {et~al.}(2019){Giacintucci}, {Markevitch}, {Cassano},
  {Venturi}, {Clarke}, {Kale}, \& {Cuciti}}]{giacintucci.etal.2019}
{Giacintucci}, S., {Markevitch}, M., {Cassano}, R., {et~al.} 2019, \apj, 880,
  70

\bibitem[{{Giodini} {et~al.}(2013){Giodini}, {Lovisari}, {Pointecouteau},
  {Ettori}, {Reiprich}, \& {Hoekstra}}]{giodini.etal.2013}
{Giodini}, S., {Lovisari}, L., {Pointecouteau}, E., {et~al.} 2013, \ssr, 177,
  247

\bibitem[{{Heinrich} {et~al.}(2021){Heinrich}, {Chen}, {Heinz}, {Zhuravleva},
  \& {Churazov}}]{heinrich.etal.2021}
{Heinrich}, A.~M., {Chen}, Y.-H., {Heinz}, S., {Zhuravleva}, I., \& {Churazov},
  E. 2021, \mnras, 505, 4646

\bibitem[{{Hillel} \& {Soker}(2017)}]{hillel.soker.2017}
{Hillel}, S. \& {Soker}, N. 2017, \mnras, 466, L39

\bibitem[{{Hitomi Collaboration} {et~al.}(2016){Hitomi Collaboration},
  {Aharonian}, {Akamatsu}, {Akimoto}, {Allen}, {Anabuki}, {Angelini}, {Arnaud},
  {Audard}, {Awaki}, {Axelsson}, {Bamba}, {Bautz}, {Blandford}, {Brenneman},
  {Brown}, {Bulbul}, {Cackett}, {Chernyakova}, {Chiao}, {Coppi}, {Costantini},
  {de Plaa}, {den Herder}, {Done}, {Dotani}, {Ebisawa}, {Eckart}, {Enoto},
  {Ezoe}, {Fabian}, {Ferrigno}, {Foster}, {Fujimoto}, {Fukazawa}, {Furuzawa},
  {Galeazzi}, {Gallo}, {Gandhi}, {Giustini}, {Goldwurm}, {Gu}, {Guainazzi},
  {Haba}, {Hagino}, {Hamaguchi}, {Harrus}, {Hatsukade}, {Hayashi}, {Hayashi},
  {Hayashida}, {Hiraga}, {Hornschemeier}, {Hoshino}, {Hughes}, {Iizuka},
  {Inoue}, {Inoue}, {Ishibashi}, {Ishida}, {Ishikawa}, {Ishisaki}, {Itoh},
  {Iyomoto}, {Kaastra}, {Kallman}, {Kamae}, {Kara}, {Kataoka}, {Katsuda},
  {Katsuta}, {Kawaharada}, {Kawai}, {Kelley}, {Khangulyan}, {Kilbourne},
  {King}, {Kitaguchi}, {Kitamoto}, {Kitayama}, {Kohmura}, {Kokubun}, {Koyama},
  {Koyama}, {Kretschmar}, {Krimm}, {Kubota}, {Kunieda}, {Laurent}, {Lebrun},
  {Lee}, {Leutenegger}, {Limousin}, {Loewenstein}, {Long}, {Lumb}, {Madejski},
  {Maeda}, {Maier}, {Makishima}, {Markevitch}, {Matsumoto}, {Matsushita},
  {McCammon}, {McNamara}, {Mehdipour}, {Miller}, {Miller}, {Mineshige},
  {Mitsuda}, {Mitsuishi}, {Miyazawa}, {Mizuno}, {Mori}, {Mori}, {Moseley},
  {Mukai}, {Murakami}, {Murakami}, {Mushotzky}, {Nagino}, {Nakagawa},
  {Nakajima}, {Nakamori}, {Nakano}, {Nakashima}, {Nakazawa}, {Nobukawa},
  {Noda}, {Nomachi}, {O'Dell}, {Odaka}, {Ohashi}, {Ohno}, {Okajima}, {Ota},
  {Ozaki}, {Paerels}, {Paltani}, {Parmar}, {Petre}, {Pinto}, {Pohl}, {Porter},
  {Pottschmidt}, {Ramsey}, {Reynolds}, {Russell}, {Safi-Harb}, {Saito},
  {Sakai}, {Sameshima}, {Sato}, {Sato}, {Sato}, {Sawada}, {Schartel},
  {Serlemitsos}, {Seta}, {Shidatsu}, {Simionescu}, {Smith}, {Soong}, {Stawarz},
  {Sugawara}, {Sugita}, {Szymkowiak}, {Tajima}, {Takahashi}, {Takahashi},
  {Takeda}, {Takei}, {Tamagawa}, {Tamura}, {Tamura}, {Tanaka}, {Tanaka},
  {Tanaka}, {Tashiro}, {Tawara}, {Terada}, {Terashima}, {Tombesi}, {Tomida},
  {Tsuboi}, {Tsujimoto}, {Tsunemi}, {Tsuru}, {Uchida}, {Uchiyama}, {Uchiyama},
  {Ueda}, {Ueda}, {Ueno}, {Uno}, {Urry}, {Ursino}, {de Vries}, {Watanabe},
  {Werner}, {Wik}, {Wilkins}, {Williams}, {Yamada}, {Yamaguchi}, {Yamaoka},
  {Yamasaki}, {Yamauchi}, {Yamauchi}, {Yaqoob}, {Yatsu}, {Yonetoku}, {Yoshida},
  {Yuasa}, {Zhuravleva}, \& {Zoghbi}}]{hitomi.collaboration.2016}
{Hitomi Collaboration}, {Aharonian}, F., {Akamatsu}, H., {et~al.} 2016, \nat,
  535, 117

\bibitem[{{Hitomi Collaboration} {et~al.}(2018){Hitomi Collaboration},
  {Aharonian}, {Akamatsu}, {Akimoto}, {Allen}, {Angelini}, {Audard}, {Awaki},
  {Axelsson}, {Bamba}, {Bautz}, {Blandford}, {Brenneman}, {Brown}, {Bulbul},
  {Cackett}, {Chernyakova}, {Chiao}, {Coppi}, {Costantini}, {de Plaa}, {de
  Vries}, {den Herder}, {Done}, {Dotani}, {Ebisawa}, {Eckart}, {Enoto}, {Ezoe},
  {Fabian}, {Ferrigno}, {Foster}, {Fujimoto}, {Fukazawa}, {Furukawa},
  {Furuzawa}, {Galeazzi}, {Gallo}, {Gandhi}, {Giustini}, {Goldwurm}, {Gu},
  {Guainazzi}, {Haba}, {Hagino}, {Hamaguchi}, {Harrus}, {Hatsukade}, {Hayashi},
  {Hayashi}, {Hayashida}, {Hiraga}, {Hornschemeier}, {Hoshino}, {Hughes},
  {Ichinohe}, {Iizuka}, {Inoue}, {Inoue}, {Ishida}, {Ishikawa}, {Ishisaki},
  {Iwai}, {Kaastra}, {Kallman}, {Kamae}, {Kataoka}, {Katsuda}, {Kawai},
  {Kelley}, {Kilbourne}, {Kitaguchi}, {Kitamoto}, {Kitayama}, {Kohmura},
  {Kokubun}, {Koyama}, {Koyama}, {Kretschmar}, {Krimm}, {Kubota}, {Kunieda},
  {Laurent}, {Lee}, {Leutenegger}, {Limousin}, {Loewenstein}, {Long}, {Lumb},
  {Madejski}, {Maeda}, {Maier}, {Makishima}, {Markevitch}, {Matsumoto},
  {Matsushita}, {McCammon}, {McNamara}, {Mehdipour}, {Miller}, {Miller},
  {Mineshige}, {Mitsuda}, {Mitsuishi}, {Miyazawa}, {Mizuno}, {Mori}, {Mori},
  {Mukai}, {Murakami}, {Mushotzky}, {Nakagawa}, {Nakajima}, {Nakamori},
  {Nakashima}, {Nakazawa}, {Nobukawa}, {Nobukawa}, {Noda}, {Odaka},
  {Ogorzalek}, {Ohashi}, {Ohno}, {Okajima}, {Ota}, {Ozaki}, {Paerels},
  {Paltani}, {Petre}, {Pinto}, {Porter}, {Pottschmidt}, {Reynolds},
  {Safi-Harb}, {Saito}, {Sakai}, {Sasaki}, {Sato}, {Sato}, {Sato}, {Sawada},
  {Schartel}, {Serlemtsos}, {Seta}, {Shidatsu}, {Simionescu}, {Smith}, {Soong},
  {Stawarz}, {Sugawara}, {Sugita}, {Szymkowiak}, {Tajima}, {Takahashi},
  {Takahashi}, {Takeda}, {Takei}, {Tamagawa}, {Tamura}, {Tanaka}, {Tanaka},
  {Tanaka}, {Tashiro}, {Tawara}, {Terada}, {Terashima}, {Tombesi}, {Tomida},
  {Tsuboi}, {Tsujimoto}, {Tsunemi}, {Tsuru}, {Uchida}, {Uchiyama}, {Uchiyama},
  {Ueda}, {Ueda}, {Uno}, {Urry}, {Ursino}, {Watanabe}, {Werner}, {Wilkins},
  {Williams}, {Yamada}, {Yamaguchi}, {Yamaoka}, {Yamasaki}, {Yamauchi},
  {Yamauchi}, {Yaqoob}, {Yatsu}, {Yonetoku}, {Zhuravleva}, \&
  {Zoghbi}}]{hitomi.collaboration.2018}
{Hitomi Collaboration}, {Aharonian}, F., {Akamatsu}, H., {et~al.} 2018, \pasj,
  70, 10

\bibitem[{{Hlavacek-Larrondo} {et~al.}(2022){Hlavacek-Larrondo}, {Li}, \&
  {Churazov}}]{hlavacek-Larrondo.etal.2022}
{Hlavacek-Larrondo}, J., {Li}, Y., \& {Churazov}, E. 2022, {AGN Feedback in
  Groups and Clusters of Galaxies}

\bibitem[{{Kaiser}(1986)}]{kaiser.1986}
{Kaiser}, N. 1986, \mnras, 222, 323

\bibitem[{{Kravtsov} \& {Borgani}(2012)}]{kravtsov.borgani.2012}
{Kravtsov}, A.~V. \& {Borgani}, S. 2012, \araa, 50, 353

\bibitem[{{Lau} {et~al.}(2017){Lau}, {Gaspari}, {Nagai}, \&
  {Coppi}}]{lau.etal.2017}
{Lau}, E.~T., {Gaspari}, M., {Nagai}, D., \& {Coppi}, P. 2017, \apj, 849, 54

\bibitem[{{Le Brun} {et~al.}(2014){Le Brun}, {McCarthy}, {Schaye}, \&
  {Ponman}}]{lebrun.etal.2014}
{Le Brun}, A. M.~C., {McCarthy}, I.~G., {Schaye}, J., \& {Ponman}, T.~J. 2014,
  \mnras, 441, 1270

\bibitem[{{Lee} {et~al.}(2023){Lee}, {Pillepich}, {ZuHone}, {Nelson}, {Jee},
  {Nagai}, \& {Finner}}]{lee.etal.2023}
{Lee}, W., {Pillepich}, A., {ZuHone}, J., {et~al.} 2023, arXiv e-prints,
  arXiv:2311.06340

\bibitem[{{Lehle} {et~al.}(2023){Lehle}, {Nelson}, {Pillepich}, {Truong}, \&
  {Rohr}}]{lehle.etal.2023}
{Lehle}, K., {Nelson}, D., {Pillepich}, A., {Truong}, N., \& {Rohr}, E. 2023,
  arXiv e-prints, arXiv:2311.06333

\bibitem[{{Li} {et~al.}(2015){Li}, {Bryan}, {Ruszkowski}, {Voit}, {O'Shea}, \&
  {Donahue}}]{li.etal.2015}
{Li}, Y., {Bryan}, G.~L., {Ruszkowski}, M., {et~al.} 2015, \apj, 811, 73

\bibitem[{{Marinacci} {et~al.}(2018){Marinacci}, {Vogelsberger}, {Pakmor},
  {Torrey}, {Springel}, {Hernquist}, {Nelson}, {Weinberger}, {Pillepich},
  {Naiman}, \& {Genel}}]{marinacci.etal.2018}
{Marinacci}, F., {Vogelsberger}, M., {Pakmor}, R., {et~al.} 2018, \mnras, 480,
  5113

\bibitem[{{McDonald} {et~al.}(2013){McDonald}, {Benson}, {Vikhlinin},
  {Stalder}, {Bleem}, {de Haan}, {Lin}, {Aird}, {Ashby}, {Bautz}, {Bayliss},
  {Bocquet}, {Brodwin}, {Carlstrom}, {Chang}, {Cho}, {Clocchiatti}, {Crawford},
  {Crites}, {Desai}, {Dobbs}, {Dudley}, {Foley}, {Forman}, {George},
  {Gettings}, {Gladders}, {Gonzalez}, {Halverson}, {High}, {Holder},
  {Holzapfel}, {Hoover}, {Hrubes}, {Jones}, {Joy}, {Keisler}, {Knox}, {Lee},
  {Leitch}, {Liu}, {Lueker}, {Luong-Van}, {Mantz}, {Marrone}, {McMahon},
  {Mehl}, {Meyer}, {Miller}, {Mocanu}, {Mohr}, {Montroy}, {Murray},
  {Nurgaliev}, {Padin}, {Plagge}, {Pryke}, {Reichardt}, {Rest}, {Ruel}, {Ruhl},
  {Saliwanchik}, {Saro}, {Sayre}, {Schaffer}, {Shirokoff}, {Song},
  {{\v{S}}uhada}, {Spieler}, {Stanford}, {Staniszewski}, {Stark}, {Story}, {van
  Engelen}, {Vanderlinde}, {Vieira}, {Williamson}, {Zahn}, \&
  {Zenteno}}]{McDonald.etal.2013}
{McDonald}, M., {Benson}, B.~A., {Vikhlinin}, A., {et~al.} 2013, \apj, 774, 23

\bibitem[{{McNamara} {et~al.}(2000){McNamara}, {Wise}, {Nulsen}, {David},
  {Sarazin}, {Bautz}, {Markevitch}, {Vikhlinin}, {Forman}, {Jones}, \&
  {Harris}}]{McNamara.etal.2000}
{McNamara}, B.~R., {Wise}, M., {Nulsen}, P.~E.~J., {et~al.} 2000, \apjl, 534,
  L135

\bibitem[{{Naiman} {et~al.}(2018){Naiman}, {Pillepich}, {Springel},
  {Ramirez-Ruiz}, {Torrey}, {Vogelsberger}, {Pakmor}, {Nelson}, {Marinacci},
  {Hernquist}, {Weinberger}, \& {Genel}}]{naiman.etal.2018}
{Naiman}, J.~P., {Pillepich}, A., {Springel}, V., {et~al.} 2018, \mnras, 477,
  1206

\bibitem[{{Nelson} {et~al.}(2023){Nelson}, {Pillepich}, {Ayromlou}, {Lee},
  {Lehle}, {Rohr}, \& {Truong}}]{nelson.etal.2023}
{Nelson}, D., {Pillepich}, A., {Ayromlou}, M., {et~al.} 2023, arXiv e-prints,
  arXiv:2311.06338

\bibitem[{{Nelson} {et~al.}(2019{\natexlab{a}}){Nelson}, {Pillepich},
  {Springel}, {Pakmor}, {Weinberger}, {Genel}, {Torrey}, {Vogelsberger},
  {Marinacci}, \& {Hernquist}}]{nelson.etal.2019b}
{Nelson}, D., {Pillepich}, A., {Springel}, V., {et~al.} 2019{\natexlab{a}},
  \mnras, 490, 3234

\bibitem[{{Nelson} {et~al.}(2018){Nelson}, {Pillepich}, {Springel},
  {Weinberger}, {Hernquist}, {Pakmor}, {Genel}, {Torrey}, {Vogelsberger},
  {Kauffmann}, {Marinacci}, \& {Naiman}}]{nelson.etal.2018}
{Nelson}, D., {Pillepich}, A., {Springel}, V., {et~al.} 2018, \mnras, 475, 624

\bibitem[{{Nelson} {et~al.}(2019{\natexlab{b}}){Nelson}, {Springel},
  {Pillepich}, {Rodriguez-Gomez}, {Torrey}, {Genel}, {Vogelsberger}, {Pakmor},
  {Marinacci}, {Weinberger}, {Kelley}, {Lovell}, {Diemer}, \&
  {Hernquist}}]{nelson.etal.2019a}
{Nelson}, D., {Springel}, V., {Pillepich}, A., {et~al.} 2019{\natexlab{b}},
  Computational Astrophysics and Cosmology, 6, 2

\bibitem[{{O'Dea} {et~al.}(2008){O'Dea}, {Baum}, {Privon}, {Noel-Storr},
  {Quillen}, {Zufelt}, {Park}, {Edge}, {Russell}, {Fabian}, {Donahue},
  {Sarazin}, {McNamara}, {Bregman}, \& {Egami}}]{ODeal.etal.2008}
{O'Dea}, C.~P., {Baum}, S.~A., {Privon}, G., {et~al.} 2008, \apj, 681, 1035

\bibitem[{{P{\'e}roux} {et~al.}(2020){P{\'e}roux}, {Nelson}, {van de Voort},
  {Pillepich}, {Marinacci}, {Vogelsberger}, \& {Hernquist}}]{peroux.etal.2020}
{P{\'e}roux}, C., {Nelson}, D., {van de Voort}, F., {et~al.} 2020, \mnras, 499,
  2462

\bibitem[{{Pillepich} {et~al.}(2018{\natexlab{a}}){Pillepich}, {Nelson},
  {Hernquist}, {Springel}, {Pakmor}, {Torrey}, {Weinberger}, {Genel}, {Naiman},
  {Marinacci}, \& {Vogelsberger}}]{pillepich.etal.2018b}
{Pillepich}, A., {Nelson}, D., {Hernquist}, L., {et~al.} 2018{\natexlab{a}},
  \mnras, 475, 648

\bibitem[{{Pillepich} {et~al.}(2021){Pillepich}, {Nelson}, {Truong},
  {Weinberger}, {Martin-Navarro}, {Springel}, {Faber}, \&
  {Hernquist}}]{pillepich.etal.2021}
{Pillepich}, A., {Nelson}, D., {Truong}, N., {et~al.} 2021, arXiv e-prints,
  arXiv:2105.08062

\bibitem[{{Pillepich} {et~al.}(2018{\natexlab{b}}){Pillepich}, {Springel},
  {Nelson}, {Genel}, {Naiman}, {Pakmor}, {Hernquist}, {Torrey}, {Vogelsberger},
  {Weinberger}, \& {Marinacci}}]{pillepich.etal.2018}
{Pillepich}, A., {Springel}, V., {Nelson}, D., {et~al.} 2018{\natexlab{b}},
  \mnras, 473, 4077

\bibitem[{{Planck Collaboration} {et~al.}(2016){Planck Collaboration}, {Ade},
  {Aghanim}, {Arnaud}, {Ashdown}, {Aumont}, {Baccigalupi}, {Banday},
  {Barreiro}, {Bartlett}, {Bartolo}, {Battaner}, {Battye}, {Benabed},
  {Beno{\^\i}t}, {Benoit-L{\'e}vy}, {Bernard}, {Bersanelli}, {Bielewicz},
  {Bock}, {Bonaldi}, {Bonavera}, {Bond}, {Borrill}, {Bouchet}, {Boulanger},
  {Bucher}, {Burigana}, {Butler}, {Calabrese}, {Cardoso}, {Catalano},
  {Challinor}, {Chamballu}, {Chary}, {Chiang}, {Chluba}, {Christensen},
  {Church}, {Clements}, {Colombi}, {Colombo}, {Combet}, {Coulais}, {Crill},
  {Curto}, {Cuttaia}, {Danese}, {Davies}, {Davis}, {de Bernardis}, {de Rosa},
  {de Zotti}, {Delabrouille}, {D{\'e}sert}, {Di Valentino}, {Dickinson},
  {Diego}, {Dolag}, {Dole}, {Donzelli}, {Dor{\'e}}, {Douspis}, {Ducout},
  {Dunkley}, {Dupac}, {Efstathiou}, {Elsner}, {En{\ss}lin}, {Eriksen},
  {Farhang}, {Fergusson}, {Finelli}, {Forni}, {Frailis}, {Fraisse},
  {Franceschi}, {Frejsel}, {Galeotta}, {Galli}, {Ganga}, {Gauthier}, {Gerbino},
  {Ghosh}, {Giard}, {Giraud-H{\'e}raud}, {Giusarma}, {Gjerl{\o}w},
  {Gonz{\'a}lez-Nuevo}, {G{\'o}rski}, {Gratton}, {Gregorio}, {Gruppuso},
  {Gudmundsson}, {Hamann}, {Hansen}, {Hanson}, {Harrison}, {Helou},
  {Henrot-Versill{\'e}}, {Hern{\'a}ndez-Monteagudo}, {Herranz}, {Hildebrandt},
  {Hivon}, {Hobson}, {Holmes}, {Hornstrup}, {Hovest}, {Huang}, {Huffenberger},
  {Hurier}, {Jaffe}, {Jaffe}, {Jones}, {Juvela}, {Keih{\"a}nen}, {Keskitalo},
  {Kisner}, {Kneissl}, {Knoche}, {Knox}, {Kunz}, {Kurki-Suonio}, {Lagache},
  {L{\"a}hteenm{\"a}ki}, {Lamarre}, {Lasenby}, {Lattanzi}, {Lawrence}, {Leahy},
  {Leonardi}, {Lesgourgues}, {Levrier}, {Lewis}, {Liguori}, {Lilje},
  {Linden-V{\o}rnle}, {L{\'o}pez-Caniego}, {Lubin}, {Mac{\'\i}as-P{\'e}rez},
  {Maggio}, {Maino}, {Mandolesi}, {Mangilli}, {Marchini}, {Maris}, {Martin},
  {Martinelli}, {Mart{\'\i}nez-Gonz{\'a}lez}, {Masi}, {Matarrese}, {McGehee},
  {Meinhold}, {Melchiorri}, {Melin}, {Mendes}, {Mennella}, {Migliaccio},
  {Millea}, {Mitra}, {Miville-Desch{\^e}nes}, {Moneti}, {Montier}, {Morgante},
  {Mortlock}, {Moss}, {Munshi}, {Murphy}, {Naselsky}, {Nati}, {Natoli},
  {Netterfield}, {N{\o}rgaard-Nielsen}, {Noviello}, {Novikov}, {Novikov},
  {Oxborrow}, {Paci}, {Pagano}, {Pajot}, {Paladini}, {Paoletti}, {Partridge},
  {Pasian}, {Patanchon}, {Pearson}, {Perdereau}, {Perotto}, {Perrotta},
  {Pettorino}, {Piacentini}, {Piat}, {Pierpaoli}, {Pietrobon}, {Plaszczynski},
  {Pointecouteau}, {Polenta}, {Popa}, {Pratt}, {Pr{\'e}zeau}, {Prunet},
  {Puget}, {Rachen}, {Reach}, {Rebolo}, {Reinecke}, {Remazeilles}, {Renault},
  {Renzi}, {Ristorcelli}, {Rocha}, {Rosset}, {Rossetti}, {Roudier},
  {Rouill{\'e} d'Orfeuil}, {Rowan-Robinson}, {Rubi{\~n}o-Mart{\'\i}n},
  {Rusholme}, {Said}, {Salvatelli}, {Salvati}, {Sandri}, {Santos},
  {Savelainen}, {Savini}, {Scott}, {Seiffert}, {Serra}, {Shellard}, {Spencer},
  {Spinelli}, {Stolyarov}, {Stompor}, {Sudiwala}, {Sunyaev}, {Sutton},
  {Suur-Uski}, {Sygnet}, {Tauber}, {Terenzi}, {Toffolatti}, {Tomasi},
  {Tristram}, {Trombetti}, {Tucci}, {Tuovinen}, {T{\"u}rler}, {Umana},
  {Valenziano}, {Valiviita}, {Van Tent}, {Vielva}, {Villa}, {Wade}, {Wandelt},
  {Wehus}, {White}, {White}, {Wilkinson}, {Yvon}, {Zacchei}, \&
  {Zonca}}]{planck.collaboration.2016}
{Planck Collaboration}, {Ade}, P.~A.~R., {Aghanim}, N., {et~al.} 2016, \aap,
  594, A13

\bibitem[{{Planelles} {et~al.}(2014){Planelles}, {Borgani}, {Fabjan},
  {Killedar}, {Murante}, {Granato}, {Ragone-Figueroa}, \&
  {Dolag}}]{planelles.etal.2014}
{Planelles}, S., {Borgani}, S., {Fabjan}, D., {et~al.} 2014, \mnras, 438, 195

\bibitem[{{Pop} {et~al.}(2022){Pop}, {Hernquist}, {Nagai}, {Kannan},
  {Weinberger}, {Springel}, {Vogelsberger}, {Nelson}, {Pakmor}, {Pillepich}, \&
  {Torrey}}]{pop.etal.2022}
{Pop}, A.-R., {Hernquist}, L., {Nagai}, D., {et~al.} 2022, arXiv e-prints,
  arXiv:2205.11528

\bibitem[{{Puchwein} {et~al.}(2008){Puchwein}, {Sijacki}, \&
  {Springel}}]{puchwein.etal.2008}
{Puchwein}, E., {Sijacki}, D., \& {Springel}, V. 2008, \apjl, 687, L53

\bibitem[{{Ramesh} {et~al.}(2023){Ramesh}, {Nelson}, {Heesen}, \&
  {Br{\"u}ggen}}]{ramesh.etal.2023d}
{Ramesh}, R., {Nelson}, D., {Heesen}, V., \& {Br{\"u}ggen}, M. 2023, arXiv
  e-prints, arXiv:2305.11214

\bibitem[{{Rasia} {et~al.}(2015){Rasia}, {Borgani}, {Murante}, {Planelles},
  {Beck}, {Biffi}, {Ragone-Figueroa}, {Granato}, {Steinborn}, \&
  {Dolag}}]{rasia.etal.2015}
{Rasia}, E., {Borgani}, S., {Murante}, G., {et~al.} 2015, \apjl, 813, L17

\bibitem[{{Rohr} {et~al.}(2023){Rohr}, {Pillepich}, {Nelson}, {Ayromlou}, \&
  {Zinger}}]{rohr.etal.2023}
{Rohr}, E., {Pillepich}, A., {Nelson}, D., {Ayromlou}, M., \& {Zinger}, E.
  2023, arXiv e-prints, arXiv:2311.06337

\bibitem[{{Sanders} \& {Fabian}(2007)}]{sanders.etal.2007}
{Sanders}, J.~S. \& {Fabian}, A.~C. 2007, \mnras, 381, 1381

\bibitem[{{Sanders} {et~al.}(2004){Sanders}, {Fabian}, {Allen}, \&
  {Schmidt}}]{sanders.etal.2004}
{Sanders}, J.~S., {Fabian}, A.~C., {Allen}, S.~W., \& {Schmidt}, R.~W. 2004,
  \mnras, 349, 952

\bibitem[{{Sanders} {et~al.}(2016){Sanders}, {Fabian}, {Russell}, {Walker}, \&
  {Blundell}}]{sanders.etal.2016}
{Sanders}, J.~S., {Fabian}, A.~C., {Russell}, H.~R., {Walker}, S.~A., \&
  {Blundell}, K.~M. 2016, \mnras, 460, 1898

\bibitem[{{Short} {et~al.}(2010){Short}, {Thomas}, {Young}, {Pearce},
  {Jenkins}, \& {Muanwong}}]{short.etal.2010}
{Short}, C.~J., {Thomas}, P.~A., {Young}, O.~E., {et~al.} 2010, \mnras, 408,
  2213

\bibitem[{{Simionescu} {et~al.}(2011){Simionescu}, {Allen}, {Mantz}, {Werner},
  {Takei}, {Morris}, {Fabian}, {Sanders}, {Nulsen}, {George}, \&
  {Taylor}}]{simionescu.etal.2011}
{Simionescu}, A., {Allen}, S.~W., {Mantz}, A., {et~al.} 2011, Science, 331,
  1576

\bibitem[{{Smith} {et~al.}(2001){Smith}, {Brickhouse}, {Liedahl}, \&
  {Raymond}}]{smith.etal.2001}
{Smith}, R.~K., {Brickhouse}, N.~S., {Liedahl}, D.~A., \& {Raymond}, J.~C.
  2001, \apjl, 556, L91

\bibitem[{{Springel}(2010)}]{springel.2010}
{Springel}, V. 2010, \mnras, 401, 791

\bibitem[{{Springel} {et~al.}(2018){Springel}, {Pakmor}, {Pillepich},
  {Weinberger}, {Nelson}, {Hernquist}, {Vogelsberger}, {Genel}, {Torrey},
  {Marinacci}, \& {Naiman}}]{springel.etal.2018}
{Springel}, V., {Pakmor}, R., {Pillepich}, A., {et~al.} 2018, \mnras, 475, 676

\bibitem[{{Truong} {et~al.}(2023){Truong}, {Pillepich}, {Nelson}, {Bogd{\'a}n},
  {Schellenberger}, {Chakraborty}, {Forman}, {Kraft}, {Markevitch},
  {Ogorzalek}, {Oppenheimer}, {Sarkar}, {Veilleux}, {Vogelsberger}, {Wang},
  {Werner}, {Zhuravleva}, \& {Zuhone}}]{truong.etal.2023}
{Truong}, N., {Pillepich}, A., {Nelson}, D., {et~al.} 2023, \mnras, 525, 1976

\bibitem[{{Truong} {et~al.}(2021){Truong}, {Pillepich}, {Nelson}, {Werner}, \&
  {Hernquist}}]{truong.etal.2021b}
{Truong}, N., {Pillepich}, A., {Nelson}, D., {Werner}, N., \& {Hernquist}, L.
  2021, \mnras, 508, 1563

\bibitem[{{Truong} {et~al.}(2020){Truong}, {Pillepich}, {Werner}, {Nelson},
  {Lakhchaura}, {Weinberger}, {Springel}, {Vogelsberger}, \&
  {Hernquist}}]{truong.etal.2020}
{Truong}, N., {Pillepich}, A., {Werner}, N., {et~al.} 2020, \mnras, 494, 549

\bibitem[{{Truong} {et~al.}(2018){Truong}, {Rasia}, {Mazzotta}, {Planelles},
  {Biffi}, {Fabjan}, {Beck}, {Borgani}, {Dolag}, {Gaspari}, {Granato},
  {Murante}, {Ragone-Figueroa}, \& {Steinborn}}]{truong.etal.2018}
{Truong}, N., {Rasia}, E., {Mazzotta}, P., {et~al.} 2018, \mnras, 474, 4089

\bibitem[{{Vikhlinin} {et~al.}(2009){Vikhlinin}, {Kravtsov}, {Burenin},
  {Ebeling}, {Forman}, {Hornstrup}, {Jones}, {Murray}, {Nagai}, {Quintana}, \&
  {Voevodkin}}]{vikhlinin.etal.2009}
{Vikhlinin}, A., {Kravtsov}, A.~V., {Burenin}, R.~A., {et~al.} 2009, \apj, 692,
  1060

\bibitem[{{Vogelsberger} {et~al.}(2013){Vogelsberger}, {Genel}, {Sijacki},
  {Torrey}, {Springel}, \& {Hernquist}}]{vogelsberger.etal.2013}
{Vogelsberger}, M., {Genel}, S., {Sijacki}, D., {et~al.} 2013, \mnras, 436,
  3031

\bibitem[{{Voit}(2005)}]{voit.2005}
{Voit}, G.~M. 2005, Reviews of Modern Physics, 77, 207

\bibitem[{{Walker} {et~al.}(2018){Walker}, {Sanders}, \&
  {Fabian}}]{walker.etal.2018}
{Walker}, S.~A., {Sanders}, J.~S., \& {Fabian}, A.~C. 2018, \mnras, 481, 1718

\bibitem[{{Weinberger} {et~al.}(2017){Weinberger}, {Springel}, {Hernquist},
  {Pillepich}, {Marinacci}, {Pakmor}, {Nelson}, {Genel}, {Vogelsberger},
  {Naiman}, \& {Torrey}}]{weinberger.etal.2017}
{Weinberger}, R., {Springel}, V., {Hernquist}, L., {et~al.} 2017, \mnras, 465,
  3291

\bibitem[{{Weinberger} {et~al.}(2018){Weinberger}, {Springel}, {Pakmor},
  {Nelson}, {Genel}, {Pillepich}, {Vogelsberger}, {Marinacci}, {Naiman},
  {Torrey}, \& {Hernquist}}]{weinberger.etal.2018}
{Weinberger}, R., {Springel}, V., {Pakmor}, R., {et~al.} 2018, \mnras, 479,
  4056

\bibitem[{{Zhuravleva} {et~al.}(2018){Zhuravleva}, {Allen}, {Mantz}, \&
  {Werner}}]{zhuravleva.etal.2018}
{Zhuravleva}, I., {Allen}, S.~W., {Mantz}, A., \& {Werner}, N. 2018, \apj, 865,
  53

\bibitem[{{Zhuravleva} {et~al.}(2014){Zhuravleva}, {Churazov}, {Schekochihin},
  {Allen}, {Ar{\'e}valo}, {Fabian}, {Forman}, {Sanders}, {Simionescu},
  {Sunyaev}, {Vikhlinin}, \& {Werner}}]{zhuravleva.etal.2014}
{Zhuravleva}, I., {Churazov}, E., {Schekochihin}, A.~A., {et~al.} 2014, \nat,
  515, 85

\bibitem[{{ZuHone} {et~al.}(2014){ZuHone}, {Biffi}, {Hallman}, {Randall},
  {Foster}, \& {Schmid}}]{zuhone.etal.2014}
{ZuHone}, J.~A., {Biffi}, V., {Hallman}, E.~J., {et~al.} 2014, arXiv e-prints,
  arXiv:1407.1783

\bibitem[{{ZuHone} {et~al.}(2013){ZuHone}, {Markevitch}, {Brunetti}, \&
  {Giacintucci}}]{zuhone.etal.2013}
{ZuHone}, J.~A., {Markevitch}, M., {Brunetti}, G., \& {Giacintucci}, S. 2013,
  \apj, 762, 78

\bibitem[{{ZuHone} {et~al.}(2018){ZuHone}, {Miller}, {Bulbul}, \&
  {Zhuravleva}}]{zuhone.etal.2018}
{ZuHone}, J.~A., {Miller}, E.~D., {Bulbul}, E., \& {Zhuravleva}, I. 2018, \apj,
  853, 180

\bibitem[{{ZuHone} {et~al.}(2023){ZuHone}, {Vikhlinin}, {Tremblay}, {Randall},
  {Andrade-Santos}, \& {Bourdin}}]{zuhone.etal.2023}
{ZuHone}, J.~A., {Vikhlinin}, A., {Tremblay}, G.~R., {et~al.} 2023, {SOXS:
  Simulated Observations of X-ray Sources}, Astrophysics Source Code Library,
  record ascl:2301.024

\end{thebibliography}

\end{document}